\documentclass[fleqn,usenatbib]{mnras}

\usepackage{newtxtext,newtxmath}

\usepackage[T1]{fontenc}
\usepackage{ae,aecompl}

\usepackage{graphicx}	
\usepackage{amsmath}	
\usepackage[flushleft]{threeparttable}
\usepackage{soul}

\title[ASAS-SN Catalog of Variable Stars VI]{The ASAS-SN Catalog of Variable Stars VI: An All-Sky Sample of \textit{$\delta$ Scuti} Stars}

\author[T. Jayasinghe et al.]{T. Jayasinghe$^{1,2}$\thanks{E-mail: jayasinghearachchilage.1@osu.edu},
K. Z. Stanek$^{1,2}$,
C. S. Kochanek$^{1,2}$,
P. J. Vallely$^{1,2}$,
B. J. Shappee$^{3}$,
\newauthor 
T. W. -S. Holoien$^{4}$,
Todd A. Thompson$^{1,2,5}$,
J. L. Prieto$^{6,7}$,
O. Pejcha$^{8}$,
M. Fausnaugh$^{9}$,
\newauthor 
S. Otero$^{10}$,
N. Hurst$^{11}$,
D. Will$^{1,11}$
\\
$^{1}$Department of Astronomy, The Ohio State University, 140 West 18th Avenue, Columbus, OH 43210, USA\\
$^{2}$Center for Cosmology and Astroparticle Physics, The Ohio State University, 191 W. Woodruff Avenue, Columbus, OH 43210, USA\\
$^{3}$Institute for Astronomy, University of Hawaii, 2680 Woodlawn Drive, Honolulu, HI 96822,USA\\
$^{4}$Carnegie Observatories, 813 Santa Barbara Street, Pasadena, CA 91101, USA\\
$^{5}$Institute for Advanced Study, Princeton, NJ, 08540\\
$^{6}$N\'ucleo de Astronom\'ia de la Facultad de Ingenier\'ia y Ciencias, Universidad Diego Portales, Av. Ej\'ercito 441, Santiago, Chile\\
$^{7}$Millennium Institute of Astrophysics, Santiago, Chile\\
$^{8}$Institute of Theoretical Physics, Faculty of Mathematics and Physics, Charles University, Czech Republic\\
$^{9}$MIT Kavli Institute for Astrophysics and Space Research, 77 Massachusetts Avenue, 37-241, Cambridge, MA 02139, USA\\
$^{10}$The American Association of Variable Star Observers, 49 Bay State Road, Cambridge, MA 02138, USA\\
$^{11}$ASC Technology Services, 433 Mendenhall Laboratory 125 South Oval Mall Columbus OH, 43210, USA\\
}

\date{Accepted XXX. Received YYY; in original form ZZZ}

\pubyear{2019}

\begin{document}
\label{firstpage}
\pagerange{\pageref{firstpage}--\pageref{lastpage}}
\maketitle

\begin{abstract}
We characterize an all-sky catalog of ${\sim} 8,400$ $\delta$ Scuti variables in ASAS-SN, which includes ${\sim} 3,300$ new discoveries. Using distances from \textit{Gaia} DR2, we derive period-luminosity relationships (PLRs) for both the fundamental mode and overtone pulsators in the $W_{JK}$, $V$, Gaia DR2 $G$, $J$, $H$, $K_s$ and $W_1$ bands. We find that the overtone pulsators have a dominant overtone mode, with many sources pulsating in the second overtone or higher order modes. The fundamental mode pulsators have metallicity dependent periods, with $\log _{10} (\rm P){\sim}-1.1$ for  $\rm [Fe/H]<-0.3$  and $\log _{10} (\rm P){\sim}-0.9$ for $\rm [Fe/H]>0$ which leads to a period-dependant scale height. Stars with $\rm P>0.100\,d$ are predominantly located close to the Galactic disk ($\rm |Z|<0.5\,kpc$). The median period at a scale height of $\rm Z{\sim}0\, kpc$ also increases with the Galactocentric radius $\rm R$, from $\log _{10} (\rm P){\sim}-0.94$ for sources with $\rm R>9\, kpc$ to $\log _{10} (\rm P){\sim}-0.85$ for sources with $\rm R<7\, kpc$, which is indicative of a radial metallicity gradient. To illustrate potential applications of this all-sky catalog, we obtained 30 min cadence, image subtraction \textit{TESS} light curves for a sample of 10 fundamental mode and 10 overtone $\delta$ Scuti stars discovered by ASAS-SN. From this sample, we identified two new $\delta$ Scuti eclipsing binaries, ASASSN-V J071855.62$-$434247.3 and ASASSN-V J170344.20$-$615941.2 with short orbital periods of $\rm P_{orb}=2.6096$ d and $\rm P_{orb}=2.5347$ d respectively. 
\end{abstract}

\begin{keywords}
stars:variables -- stars:variables:Delta Scuti -- stars:binaries:eclipsing -- catalogues --surveys 
\end{keywords}
\clearpage


\section{Introduction}

$\delta$ Scuti stars are intermediate-mass variable stars with spectral types between A2 and F2 that pulsate at high frequencies ($\rm 0.02 \,d<P<0.25 \,d$) with typical amplitudes in the V-band of $\rm 3\,mmag<A<0.9\, mag$ \citep{2013AJ....145..132C,2017ampm.book.....B}. They have temperatures in the range $6900\, K<\rm T_{eff}<8900\,K$ while on the zero-age-main-sequence (ZAMS) and are located where the classical instability strip intersects the ZAMS \citep{2001A&A...366..178R, 1979PASP...91....5B}. $\delta$ Scuti stars have masses in the range $1.5M_{\odot}<M<2.5M_{\odot}$, which places them in the transition regime between low-mass stars having convective envelopes and high-mass stars having radiative envelopes and convective cores \citep{2017ampm.book.....B}. Thus, the pulsations in $\delta$ Scuti stars allow for detailed studies of stellar structure and evolution in this transition regime \citep{2018MNRAS.476.3169B}. Most $\delta$ Scuti stars tend to cross the instability strip on approximately horizontal tracks, placing them in the core hydrogen burning or early hydrogen shell burning evolutionary stages \citep{2000ASPC..210....3B,2010aste.book.....A}.

Most $\delta$ Scuti stars are multi-periodic pulsators, with both high amplitude radial pulsations and low amplitude non-radial pulsations \citep{2017ampm.book.....B}. These pulsations are driven by the $\kappa$ mechanism in the $\rm He \,II $ partial ionization zone \citep{2000ASPC..210....3B,2010aste.book.....A}. The $\kappa$ mechanism can produce pressure (p) modes with periods between 15 minutes and 8 hours \citep{2011A&A...534A.125U,2014MNRAS.439.2078H}. Recently, studies of \textit{Kepler} light curves found that ${\sim}25\%$ of the $\delta$ Scuti stars were in fact hybrid pulsators, having both p modes excited by the $\kappa$ mechanism and gravity (g) modes excited by the convective flux blocking mechanism \citep{2011A&A...534A.125U,2011MNRAS.417..591B,2018MNRAS.476.3169B}. The rich range of pulsation modes in $\delta$ Scuti stars have been studied extensively using both ground-based and space-based observatories \citep{1998A&A...331..271B,1999A&A...349..225B,2000ASPC..210....3B,2018MNRAS.476.3169B,2019FrASS...6...40G}.

Much like RR Lyrae stars and Cepheids that also pulsate due to the $\kappa$ mechanism, $\delta$ Scuti stars are known to follow a period-luminosity relationship \citep{1975ApJ...200..343B,1997PASP..109.1221M,2010AcA....60....1P,2019MNRAS.486.4348Z}. The scatter in the $\delta$ Scuti period-luminosity relationship (PLR) tends to be somewhat larger than that observed in the PLRs of RR Lyrae and Cepheids. $\delta$ Scuti stars can pulsate in the fundamental mode ($n=1$) and overtone modes ($n=2,3,4,...$). These modes are either radial ($l=0$) or non-radial ($l=1,2,3,4,...$) \citep{2019MNRAS.486.4348Z}. In general, the most likely pulsation mode observed in the $\delta$ Scuti PLR is the radial fundamental mode ($n=1$, $l=0$).

The All-Sky Automated Survey for SuperNovae (ASAS-SN, \citealt{2014ApJ...788...48S, 2017PASP..129j4502K}), using two units in Chile and Hawaii each with 4 telescopes, monitored the visible sky to a depth of $V\lesssim17$ mag with a cadence of 2-3 days. ASAS-SN has since expanded to 5 units with 20 telescopes and is currently monitoring the sky in the g-band to a depth of $g\lesssim18.5$ mag with a cadence of $\sim1$ day. The ASAS-SN telescopes are hosted by the Las Cumbres Observatory (LCO; \citealt{2013PASP..125.1031B}) in Hawaii, Chile, Texas and South Africa. ASAS-SN is well suited for the characterization of stellar variability across the whole sky due to its excellent baseline and all-sky coverage. 

In a series of papers \citep{2018MNRAS.477.3145J,2019MNRAS.485..961J,2019arXiv190710609J}, we have been systematically identifying and classifying variables from the ASAS-SN V-band data. This includes discovering  ${\sim}220,000$ new variables and homogeneously classifying both the new and previously known variables in this magnitude range \citep{2019MNRAS.486.1907J}. We are also exploring the synergies between ASAS-SN and large-scale spectroscopic surveys starting with APOGEE \citep{2015AJ....150..148H,2019MNRAS.tmp.1644P,2018arXiv180602751T}.

Here, we analyze an all-sky catalogue of 8418 $\delta$ Scuti stars in the ASAS-SN V-band data. In Section $\S2$, we summarize the ASAS-SN catalogue of $\delta$ Scuti stars and the cross-matching to external photometric and spectroscopic catalogues. We derive period luminosity relations (PLRs) in Section $\S3$ and analyze the sample of $\delta$ Scuti stars with spectroscopic cross-matches in Section $\S4$. We present an analysis of 20 $\delta$ Scuti stars discovered by ASAS-SN using \textit{TESS} light curves in Section $\S5$ and summarize our work in Section $\S6$. The V-band light curves and other variability and photometric information for all of the ${\sim}8,400$ sources studied in this work are available online at the ASAS-SN variable stars database (\url{https://asas-sn.osu.edu/variables}). 

\section{The ASAS-SN Catalog of $\delta$ Scuti stars}

In this work, we selected 8418 $\delta$ Scuti stars identified during our systematic search for variables. This includes new $\delta$ Scuti stars in the Northern hemisphere (Jayasinghe et al. 2019, in prep) and regions of the southern Galactic plane that were missed in the previous survey papers \citep{2019arXiv190710609J}. Out of the 8418 $\delta$ Scuti stars in this catalogue, 3322 (${\sim}40\%$) are new ASAS-SN discoveries. The ASAS-SN V-band observations used in this work were made by the ``Brutus" (Haleakala, Hawaii) and ``Cassius" (CTIO, Chile) quadruple telescopes between 2013 and 2018. Each ASAS-SN V-band field is observed to a depth of $V\lesssim17$ mag. The field of view of an ASAS-SN camera is 4.5 deg$^2$, the pixel scale is 8\farcs0 and the FWHM is typically ${\sim}2$ pixels. ASAS-SN tends to saturate at ${\sim} 10-11$ mag, but we attempt to correct the light curves of saturated sources for bleed trails (see \citealt{2017PASP..129j4502K}). The V-band light curves were extracted as described in \citet{2018MNRAS.477.3145J} using image subtraction \citep{1998ApJ...503..325A,2000A&AS..144..363A} and aperture photometry on the subtracted images with a 2 pixel radius aperture. The APASS catalog \citep{2015AAS...22533616H} and the ATLAS All-Sky Stellar Reference Catalog \citep{2018ApJ...867..105T} were used for calibration. We corrected the zero point offsets between the different cameras as described in \citet{2018MNRAS.477.3145J}. The photometric errors were recalculated as described in \citet{2019MNRAS.485..961J}. Variable sources were identified and subsequently classified using two independent random forest classifiers and quality checks as described in \citet{2019MNRAS.486.1907J,2019arXiv190710609J}. We used the \verb"astropy" implementation of the Generalized Lomb-Scargle (GLS, \citealt{2009A&A...496..577Z,1982ApJ...263..835S}) periodogram to search for periodicity over the range $0.05\leq P \leq1000$ days for these $\delta$ Scuti stars.

We cross-matched the $\delta$ Scuti stars with Gaia DR2 \citep{2018arXiv180409365G} using a matching radius of 5\farcs0. The sources were assigned distance estimates from the Gaia DR2 probabilistic distance estimates \citep{2018AJ....156...58B} by cross-matching based on the Gaia DR2 \verb"source_id". We also cross-matched these sources to the 2MASS \citep{2006AJ....131.1163S} and AllWISE \citep{2013yCat.2328....0C,2010AJ....140.1868W} catalogues using a matching radius of 10\farcs0. We used \verb"TOPCAT" \citep{2005ASPC..347...29T} for this process. Following the cross-matching process, we calculated the absolute, reddening-free Wesenheit magnitude \citep{1982ApJ...253..575M,2018arXiv180803659L} for each source as 
\begin{equation}
    W_{JK}=M_{\rm K_s}-0.686(J-K_s) \,.
	\label{eq:wk}
\end{equation} For each source, we also calculate the total line of sight Galactic reddening $E(B-V)$ from the recalibrated `SFD' dust maps \citep{2011ApJ...737..103S,1998ApJ...500..525S} using \verb"dustmaps" \citep{2018JOSS....3..695G}.

To explore the synergy between ASAS-SN and large spectroscopic surveys, we also cross-matched the variables with the APOGEE DR15 catalog \citep{2015AJ....150..148H, 2017AJ....154...94M}, the RAVE-on catalog \citep{2017ApJ...840...59C}, the LAMOST DR5 v4 catalog \citep{2012RAA....12.1197C} and the GALAH DR2 catalog \citep{2015MNRAS.449.2604D,2018MNRAS.478.4513B} using a matching radius of 5\farcs0. We identified 972 matches to the $\delta$ Scuti stars, with all the cross-matches coming from the LAMOST ($86.7\%$), GALAH ($10.0\%$), or RAVE ($3.2\%$) surveys. The LAMOST DR5 v4 catalog also lists the spectral type of each source, and the majority of the cross-matches from LAMOST have a spectral type of F0 ($55.6\%$), A7V ($20.1\%$) or A6IV ($5.0\%$).

\begin{figure*}
	\includegraphics[width=\textwidth]{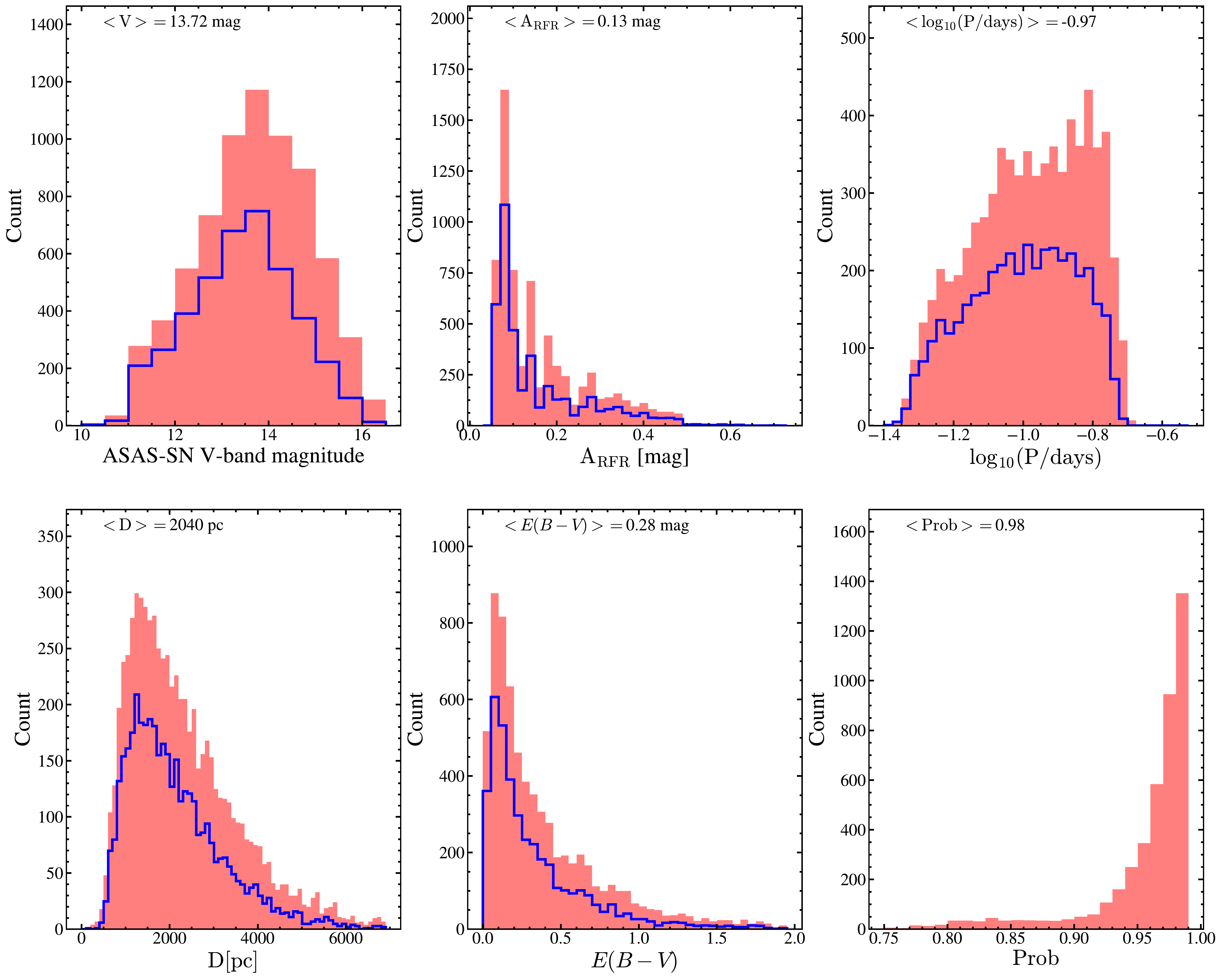}
    \caption{Distribution of the ASAS-SN $\delta$ Scuti stars in their V-band magnitude, amplitude ($\rm A_{RFR}$), period, distance, reddening and classification probability. The median value for each feature is listed. The blue histograms show the distribution of $\delta$ Scuti stars with classification probabilities greater than the median classification probability for the entire sample ($\rm Prob>0.98$).}
    \label{fig:fig1}
\end{figure*}

We illustrate the distribution of the ASAS-SN $\delta$ Scuti stars in their V-band magnitude, amplitude ($\rm A_{RFR}$), period, distance, reddening and classification probability in Figure \ref{fig:fig1}. The variability amplitude $\rm A_{RFR}$ is calculated using a non-parametric random forest regression model \citep{2019MNRAS.486.1907J}. The majority of the sources are brighter than $V=16$ mag, as ASAS-SN loses sensitivity to low-amplitude variability at fainter magnitudes. The distribution of the $\delta$ Scuti stars in amplitude is as expected, with the vast majority having amplitudes $\rm A_{RFR}<0.30$ mag. Classification probabilities of $\rm Prob>0.9$ are very reliable and ${\sim}92\%$ of our sample of $\delta$ Scuti stars have $\rm Prob>0.9$ and the median classification probability is $\rm Prob=0.98$. The distribution of these sources in period has a sharp cutoff around $P{\sim}0.2$ d ($\log _{10} (\rm P/days){\sim}-0.7$). This is an artificial cutoff implemented in our classification pipeline to minimize contamination from short period RR Lyrae and contact binary stars. The distribution of periods appears to show an excess of sources around $\log _{10} (\rm P/days){\sim}-0.8$, but this excess is not seen in the distribution of periods for the subset of sources with $\rm Prob>0.98$, possibly indicating some contamination from other variable groups.

\begin{figure*}
	\includegraphics[width=\textwidth]{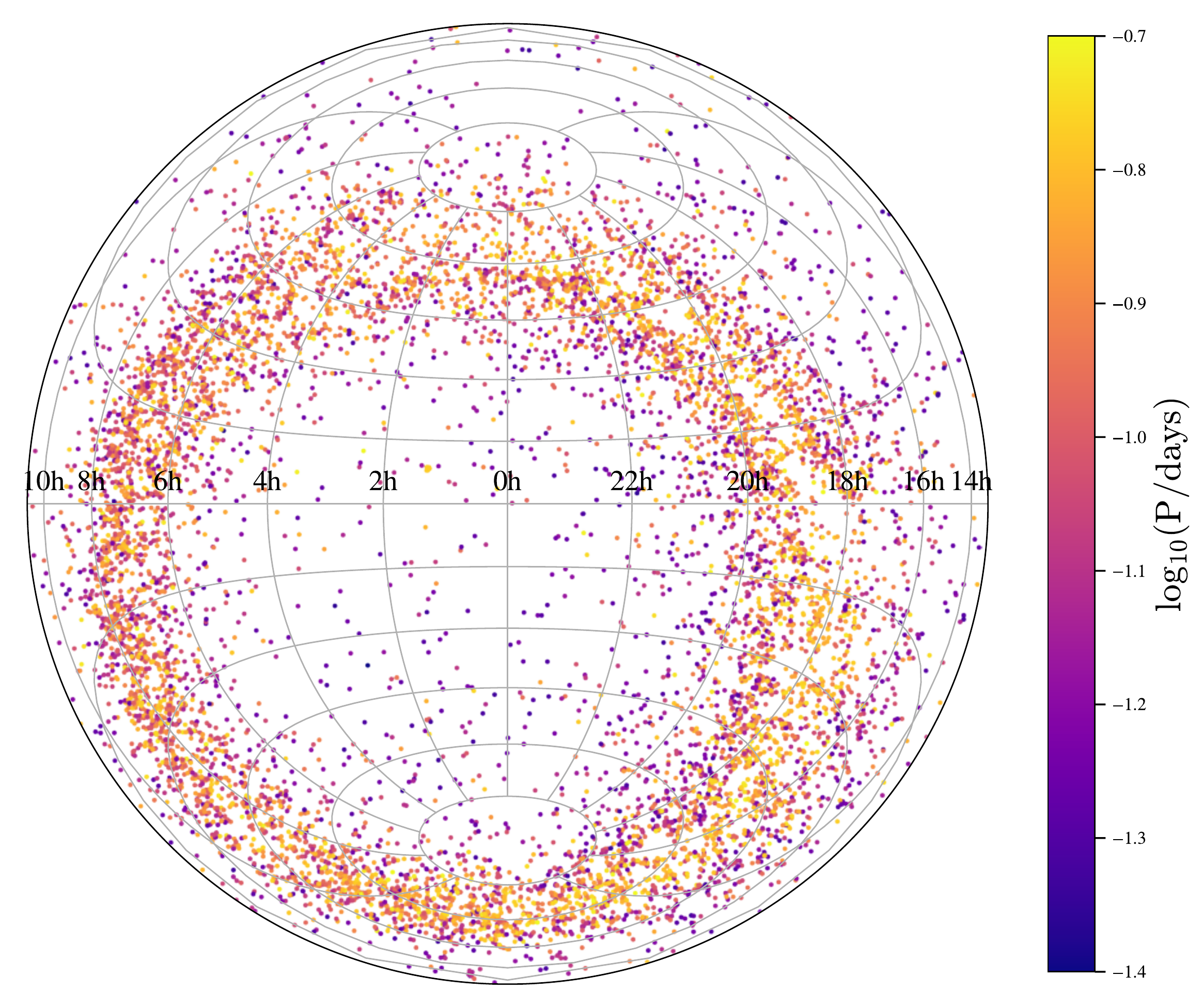}
    \caption{Spatial distribution of the $\sim 8400$ $\delta$ Scuti stars in Equatorial coordinates (Lambert projection). The points are colored by period. Note how the longer period stars tend to be closer to the Galactic plane. }
    \label{fig:fig2}
\end{figure*}

There are 746 sources within $1\, \rm kpc$ but most (${\sim}91\%$) of the sources are located further away. A large fraction have useful parallaxes, as ${\sim}79\%$ (${\sim}63\%$) of the sources have \verb"parallax"/\verb"parallax_error" $>5$ ($>10$). The median value of the reddening $E(B-V)$ indicates substantial extinction ($A_V {\sim}0.87$ mag), assuming $R_V=3.1$ dust \citep{1989ApJ...345..245C}. This is not too surprising as these stars tend to be located towards the Galactic disk with $75\%$ of the $\delta$ Scuti stars within $\rm |b|<18.4$ deg. The sky distribution of the $\delta$ Scuti stars in ASAS-SN, colored by their period, is shown in Figure \ref{fig:fig2}. There is a clear gradient of increasing period towards lower Galactic latitudes, which suggests a metallicity dependence to the period of these stars. We will explore this dependence further in Section $\S4$.

High-amplitude $\delta$ Scuti (HADS) stars are a subset of the $\delta$ Scuti class that pulsate in radial modes with large amplitudes \citep{2000ASPC..210..373M}. The classification scheme that is used by ASAS-SN follows the definitions of the VSX catalog \citep{2006SASS...25...47W}, which defines HADS stars as those with V-band amplitudes $\rm A>0.15$ mag. However, canonically, HADS stars have been defined with an amplitude cutoff of  $\rm A>0.30$ mag. With the VSX definition, our catalog contained 3989 HADS sources, of which 3660 (2086) have $\rm Prob>0.9$ ($\rm Prob>0.98$). With the canonical definition, our catalog contained 1496 HADS sources, of which 1406 (865) have $\rm Prob>0.9$ ($\rm Prob>0.98$).

\section{Period-luminosity relationships}

\begin{figure}
	\includegraphics[width=0.5\textwidth]{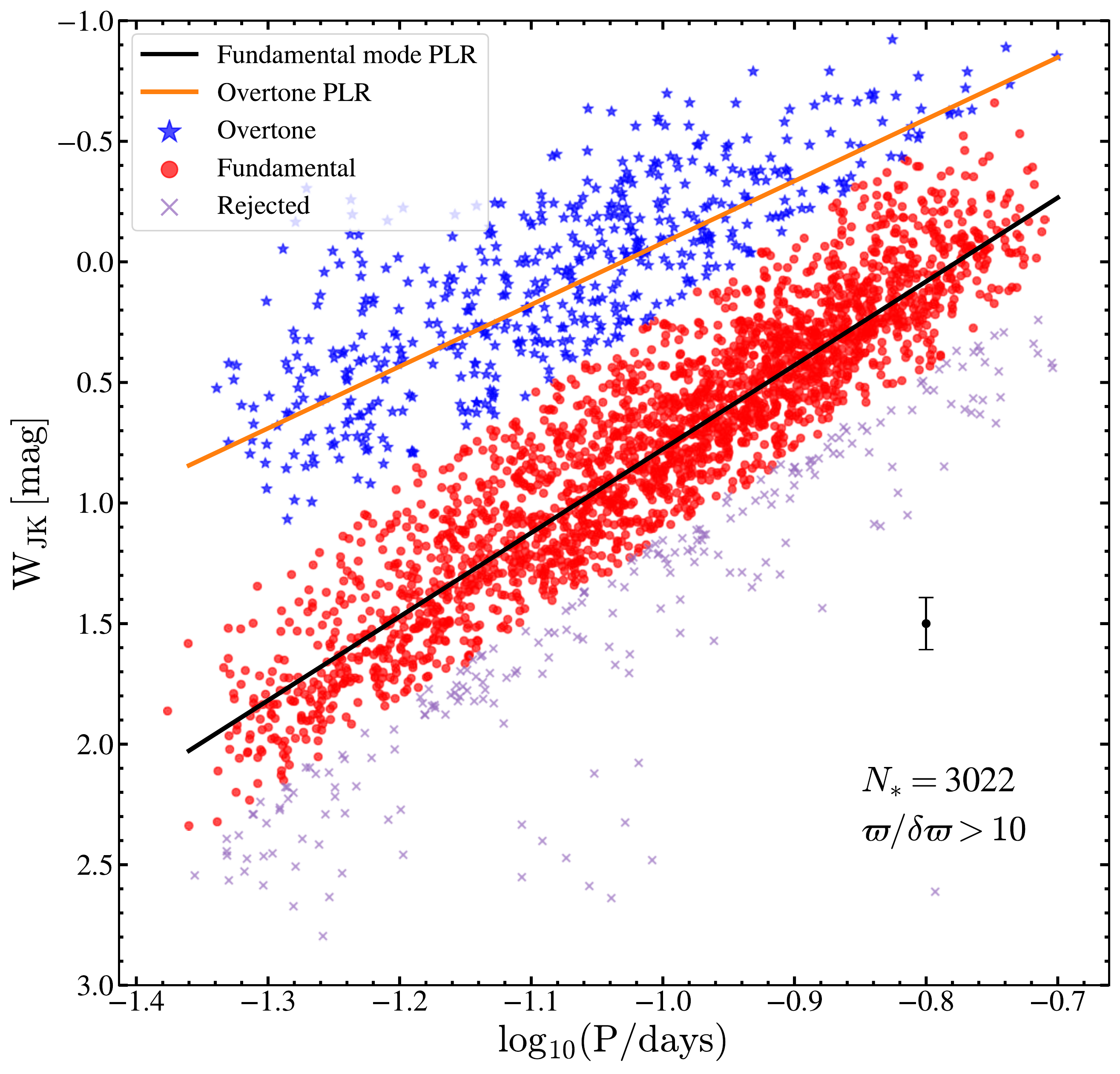}
    \caption{The Wesenheit $W_{JK}$ PLR diagram for the $\delta$ Scuti stars with $\rm Prob>0.98 $ (fundamental mode) and $\rm Prob>0.90 $ (overtone) and parallaxes better than $10\%$. The fitted PLRs for the fundamental mode and overtone pulsators are shown as black and orange lines respectively. The average uncertainties are shown in black.}
    \label{fig:fig3}
\end{figure}

\begin{figure}
	\includegraphics[width=0.5\textwidth]{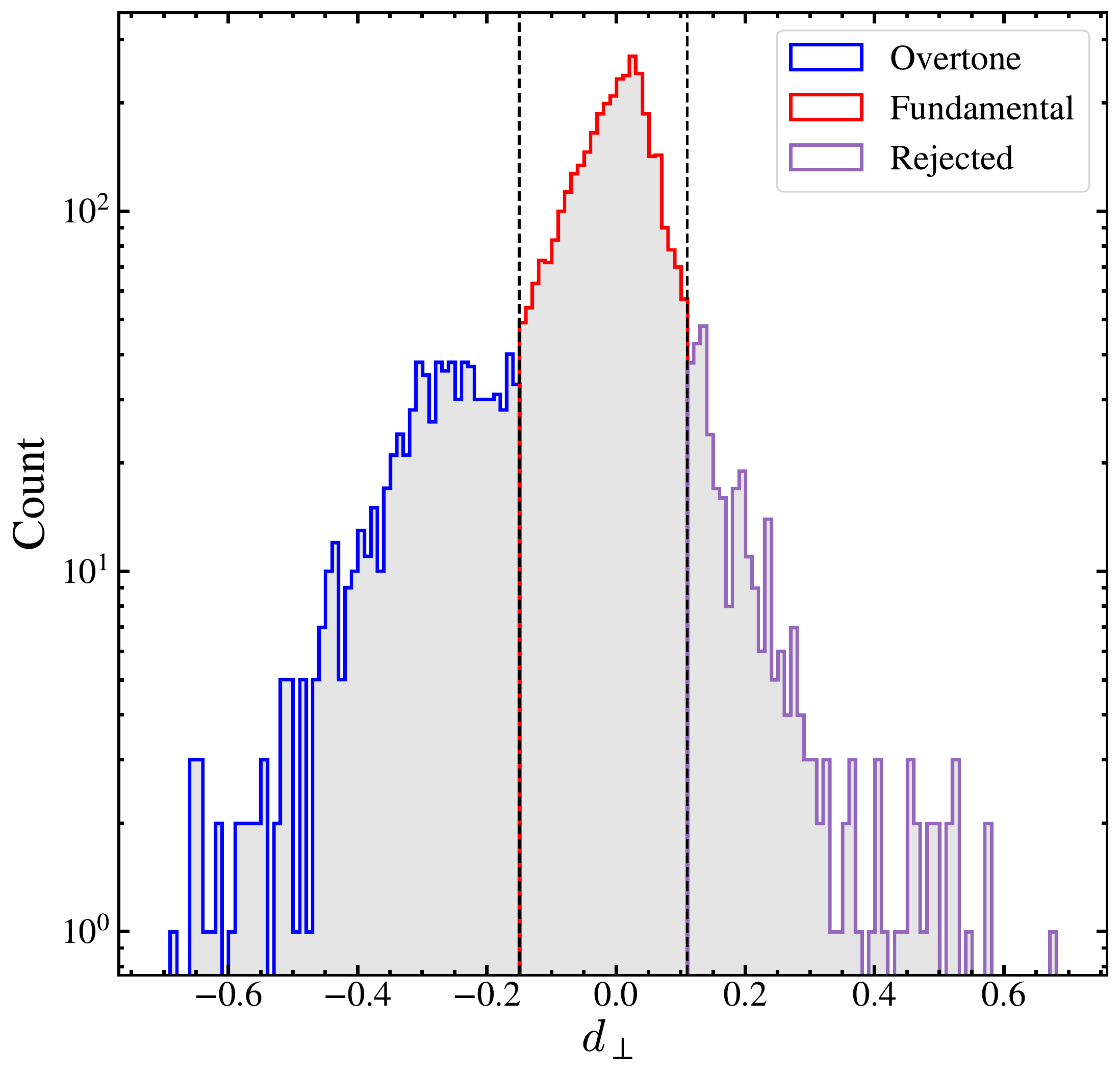}
    \caption{Distribution of the perpendicular distances to an initial fundamental mode Wesenheit $W_{JK}$ PLR fit. The distributions of the fundamental mode pulsators (red), likely overtone pulsators (blue) and rejected candidates (purple) are shown as histograms. The shaded gray histogram is for the entire sample.}
    \label{fig:fig4}
\end{figure}

Figure \ref{fig:fig3} shows the Wesenheit $W_{JK}$ period-luminosity relationship (PLR) diagram for the $\delta$ Scuti stars. We identify two possible PLRs corresponding to the fundamental and overtone modes of pulsation (Figure \ref{fig:fig4}). Previous studies have derived PLRs for the $\delta$ Scuti stars using a smaller number of sources \citep{2019MNRAS.486.4348Z,2011AJ....142..110M}. In addition, similar PLR sequences have been previously studied for other classes of pulsating variables including RR Lyrae and Cepheids (see, for e.g., the review by \citealt{2018SSRv..214..113B}). 

We sorted the sample of sources into three categories consisting of fundamental mode pulsators, likely overtone pulsators and rejected candidates. Initially, we empirically selected a sample of ${\sim}1500$ fundamental mode pulsators from our catalog and fit a PLR with the form $$W_{JK}=\rm -3.495\log_{10}(P/0.1\,d)+0.767.$$ Next, we calculated the perpendicular distances ($d_\perp$) from this initial PLR for each source in our catalogue (Figure \ref{fig:fig4}). Based on the distribution of these distances, we separate the $\delta$ Scuti stars into fundamental mode sources ($-0.15\leq d_\perp \leq 0.11$), overtone sources ($d_\perp<-0.15$) and rejected candidates ($d_\perp>0.11$). These cuts were motivated by the drop in the distribution centered at $d_\perp{\sim}0$, corresponding to the fundamental mode pulsators. Figure \ref{fig:fig3} shows the Wesenheit $W_{JK}$ PLR diagram for these sources, with the fitted PLRs shown. The set of rejected candidates are located to the right of the fundamental mode PLR sequence and likely consists of $\gamma$ Dor stars with harmonics above 5 $\rm d^{-1}$ \citep{2019MNRAS.486.4348Z,2019MNRAS.485.2380M}. Given that the overtone pulsators are rare and that the training set was biased by the large number of fundamental mode pulsators, we expect $\delta$ Scuti stars pulsating in a dominant overtone mode to have somewhat lower classification probabilities in our pipeline. Thus, we choose two different cutoffs in classification probability to improve the quality of our PLRs. When constructing PLRs, we choose fundamental mode pulsators with $\rm Prob>0.98$, which corresponds to the mean classification probability of our entire sample. For the overtone pulsators, we relax this cutoff to $\rm Prob>0.9$, which still implies a robust classification. We also implement a cutoff of \verb"parallax"/\verb"parallax_error" $>10$ to reduce the uncertainty in the absolute magnitudes. 

\begin{figure}
	\includegraphics[width=0.5\textwidth]{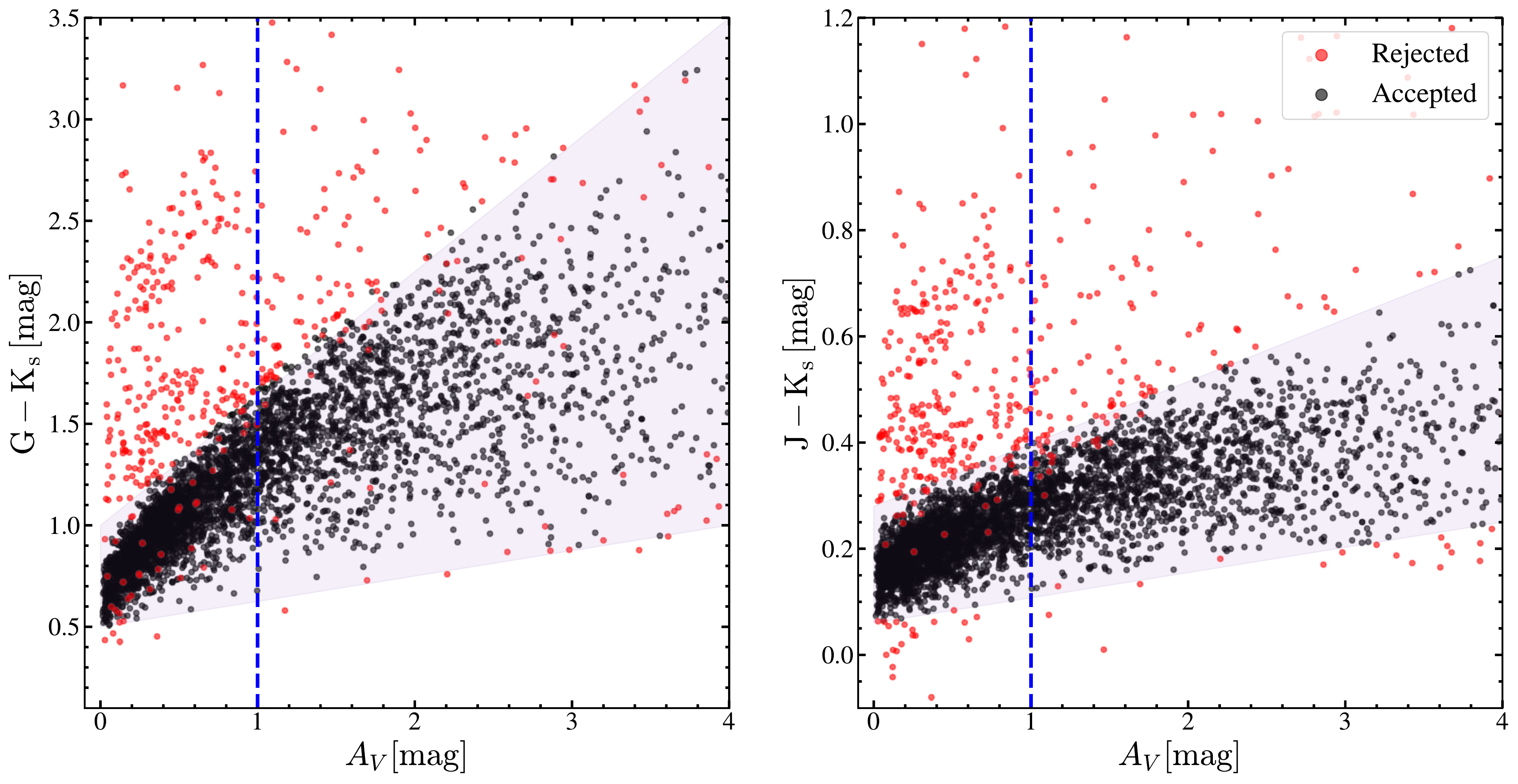}
    \caption{Distribution of the $\delta$ Scuti stars in the $G-K_s$ (left) and $J-K_s$ (right) color-extinction spaces. Sources inside (outside) the shaded region and colored in black (red) are defined to have acceptable (bad) colors and/or extinctions. The sources with bad colors and/or extinctions deviate from the expected distribution of the $\delta$ Scuti stars in the $G-K_s$ and $J-K_s$ color-extinction spaces and are not considered in our work. The curvature towards higher $A_V$ enters as the distinction between the total Galactic extinction and the actual extinction to the star becomes important.}
    \label{fig:fig5}
\end{figure}

The SFD reddening estimates are totals for the line of sight rather than for stars at a particular distance. Figure \ref{fig:fig5} shows the $G-K_s$ and $J-K_s$ colors of the stars as a function of $A_V$. These finite distance effects manifest as a curvature in the color distribution for larger $A_V$. To minimize this problem, we use a limit of $A_V<1$ mag for stars used to construct the PLRs. We also see a scattering of stars with inconsistent colors/extinctions, and we eliminate these stars as well.

For the non-Wesenheit PLRs, we corrected for the interstellar extinction using the SFD estimate of the reddening, the relative extinction coefficients $A_G/A_V=0.85926$, $A_J/A_V=0.29434$, $A_H/A_V=0.18128$, $A_{K_s}/A_V=0.11838$, $A_{W1}/A_V=0.07134$ and, \begin{equation}
M_{\lambda}=m_\lambda-5\log(d)+5-A_\lambda,
    \label{mageq}
\end{equation} where d is the Gaia DR2 probabilistic distance estimate \citep{2018AJ....156...58B}. We then fit PLRs of the form \begin{equation}
M_\lambda=\rm A\log_{10}(P/0.1\,d)+B,
    \label{plrfit}
\end{equation} using the Levenberg-Marquardt chi-square minimization routine in \verb"scikit-learn" \citep{2012arXiv1201.0490P}. After the initial fit, we removed outliers from the PLR fit for each band by calculating the distance from an initial fit $$r=\sqrt{(\Delta \log_{10} P)^2+(\Delta M_\lambda)^2},$$ where $$\Delta\log_{10}(P)=\log_{10}( P_{fit}/P_{obs})$$ and$$\Delta M_\lambda=M_{\lambda,fit}-M_{\lambda,obs}.$$ Sources that deviated from this fit by $>3\sigma_r$ were removed. After removing these outliers, the parameters from the trial fit were then used to initialize a Monte Carlo Markov Chain sampler (MCMC) with 200 walkers, that were run for 20000 iterations. We used the MCMC implementation \verb"emcee" \citep{2013PASP..125..306F}. The errors in the PLR parameters were derived from the MCMC chains.

\begin{figure*}
	\includegraphics[width=\textwidth]{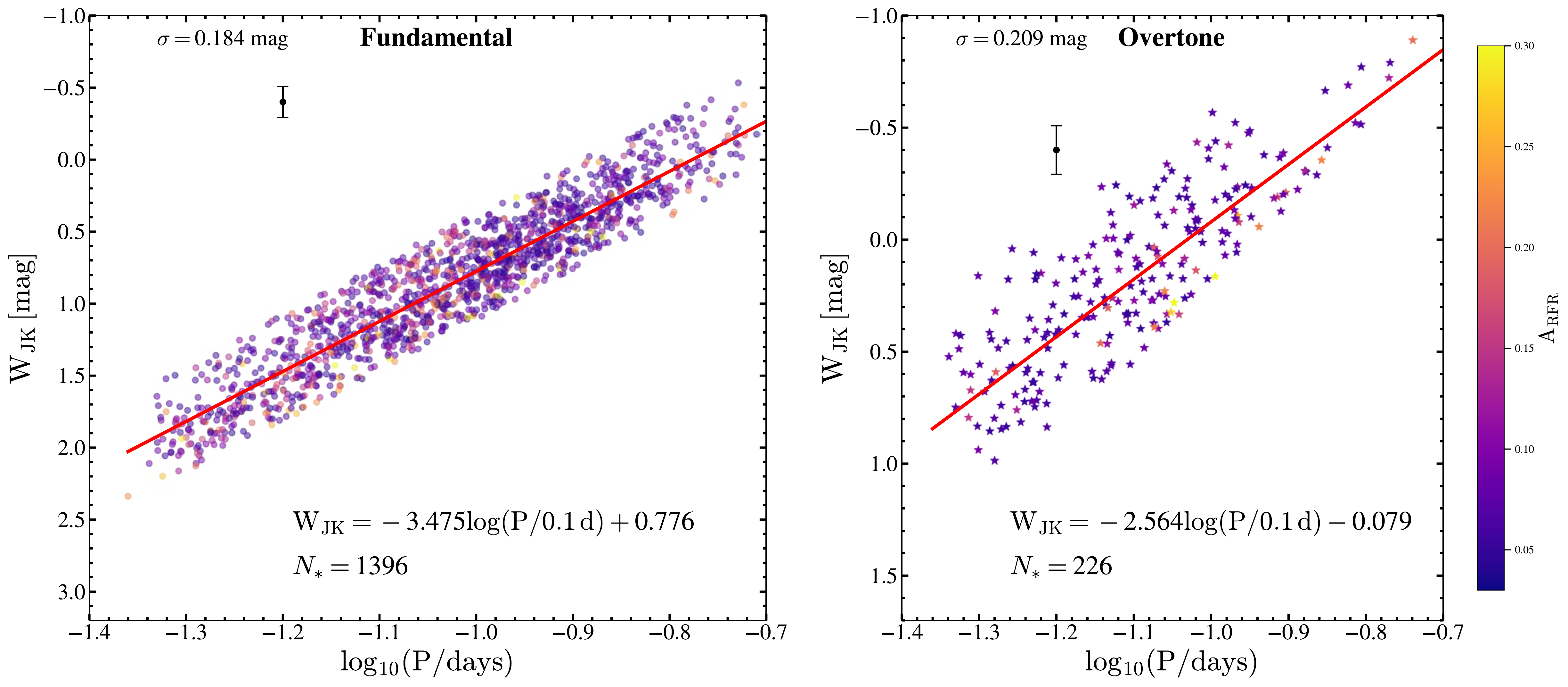}
    \caption{The Wesenheit $W_{JK}$ PLR for the fundamental mode (left) and overtone (right) $\delta$ Scuti stars. The fitted PLR is shown in red and the points are colored by the variability amplitude $\rm A_{RFR}$.}
    \label{fig:fig6}
\end{figure*}

\begin{figure*}
	\includegraphics[width=\textwidth]{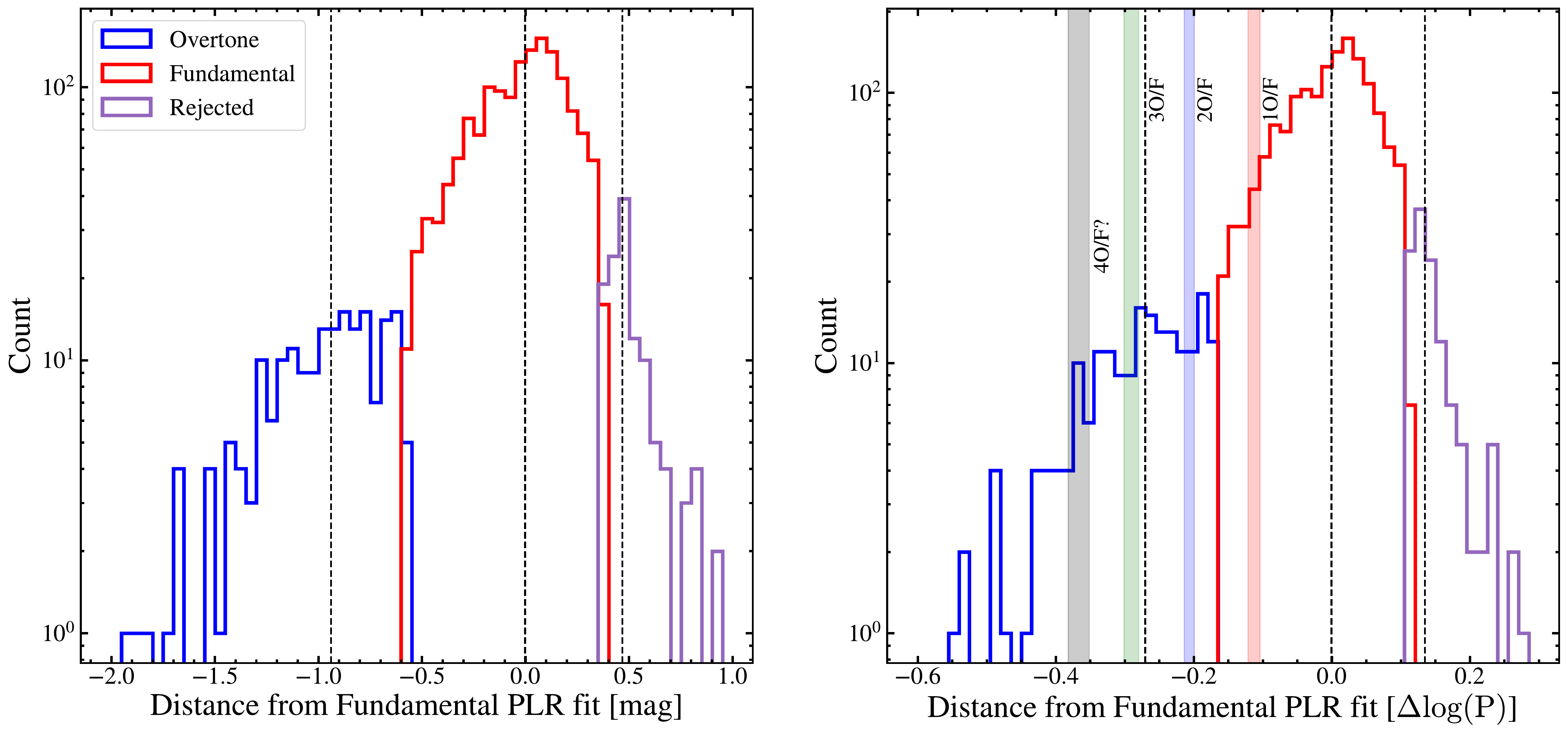}
    \caption{Distribution of the $\delta$ Scuti stars in the vertical ($\Delta W_{JK}$, left) and horizontal ($\Delta \log_{10}P$, right) deviations from the fundamental mode PLR fit. The dashed black lines show the median value of the deviations. The range of expected period ratios for $\delta$ Scuti stars pulsating in the first overtone (red), the second overtone (blue) or the third overtone (green) are shaded \citep{1979ApJ...227..935S}. The likely range of period ratios for the fourth overtone is shaded in black.}
    \label{fig:fig7}
\end{figure*}

We fit Wesenheit $W_{JK}$ magnitude PLRs to both the fundamental mode and overtone sample as shown in Figure \ref{fig:fig6}. The variability amplitudes of the overtone pulsators tend to be smaller than those of fundamental mode pulsators. The overtone sample likely encompasses sources with different dominant overtones, so the PLR fit is to an average distribution, and not a particular overtone. We also looked at the deviation from the PLR fit in the horizontal ($\Delta(\log_{10}(P))$) and vertical ($\Delta W_{JK}$) directions
(Figure \ref{fig:fig7}). Expected period ratios for combinations of the fundamental mode and the first three overtone modes are highlighted \citep{1979ApJ...227..935S}. It is clear that the distributions of these sources in $\Delta(\log_{10}P$ and $\Delta W_{JK}$ follow a bi-modal distribution, with the overtone sources clearly distinct from the fundamental mode pulsators. The overtone sources have median displacements from the fundamental mode PLR fit of $\Delta \log_{10}P{\sim}-0.27$ and $\Delta W_{JK} {\sim}-0.94$. The vertical displacement is reminiscent of overtone populations in other classical pulsators (for e.g., \citealt{1995A&A...303..137B} noted that overtone Cepheids formed a PLR sequence ${\sim}1$ mag brighter than the fundamental PLR for Cepheid variables). \citet{2019MNRAS.486.4348Z} noted an excess of sources in a ridge to the left of the fundamental ridge and found a similar displacement in $\Delta \log_{10}P$. This value of $\Delta \log_{10}P$ corresponds to a period ratio with the fundamental mode of $P/P_F{\sim}0.535$, which is consistent with the period ratios expected for the third overtone from theoretical models of $\delta$ Scuti stars \citep{1979ApJ...227..935S}. Indeed, \citet{2019MNRAS.486.4348Z} suggested the possibility of a resonant third or fourth overtone mode of pulsation as the mechanism behind the additional PLR sequence. We also note an additional peak at $\Delta \log_{10}P{\sim}-0.37$, which corresponds to a period ratio with the fundamental mode of $P_O/P_F{\sim}0.43$. We suspect that this could be the fourth overtone but we were unable to find any theoretical predictions of period ratios for the fourth overtone.

\begin{table}
	\centering
	\caption{PLR parameters for the fundamental mode $\delta$ Scuti}
	\label{tab:fundfits}
\begin{tabular}{rrrr}
		\hline
		 Band & A & B & $\sigma$\\
		  & mag & mag & mag\\		 
		\hline
		$W_{JK}$ & $-3.475 \pm 0.034$&  $0.776 \pm 0.004$ &  0.184\\
		$V$ & $-3.006 \pm 0.479$&  $1.493 \pm 0.245$ &  0.227\\
		$G$ & $-3.047 \pm 0.555$&  $1.532 \pm 0.301$ &  0.210\\
		$J$ & $-3.330 \pm 0.064$&  $0.997 \pm 0.079$ &  0.188\\	
		$H$ & $-3.373 \pm 0.037$&  $0.894 \pm 0.060$ &  0.185\\
		$K_s$ & $-3.397 \pm 0.044$&  $0.868 \pm 0.064$ &  0.183\\	
		$W_1$ & $-3.385 \pm 0.036$&  $0.834 \pm 0.036$ &  0.190\\		
\hline
\end{tabular}
\end{table}

\begin{table}
	\centering
	\caption{PLR parameters for the overtone $\delta$ Scuti}
	\label{tab:otfits}
\begin{tabular}{rrrr}
		\hline
		 Band & A & B & $\sigma$\\
		  & mag & mag & mag\\
		\hline
		$W_{JK}$ & $-2.564 \pm 0.138$&  $-0.079 \pm 0.013$ &  0.209\\
		$V$ & $-1.980 \pm 0.144$&  $0.613 \pm 0.019$ &  0.291\\
		$G$ & $-1.969 \pm 0.128$&  $0.659 \pm 0.018$ &  0.263\\
		$J$ & $-2.045 \pm 0.121$&  $0.170 \pm 0.013$ &  0.223\\	
		$H$ & $-2.299 \pm 0.147$&  $0.053 \pm 0.016$ &  0.213\\
		$K_s$ & $-2.439 \pm 0.139$&  $0.021 \pm 0.015$ &  0.209\\	
		$W_1$ & $-2.560 \pm 0.142$&  $-0.032 \pm 0.070$ &  0.218\\	
\hline
\end{tabular}
\end{table}

\begin{figure*}
	\includegraphics[width=\textwidth]{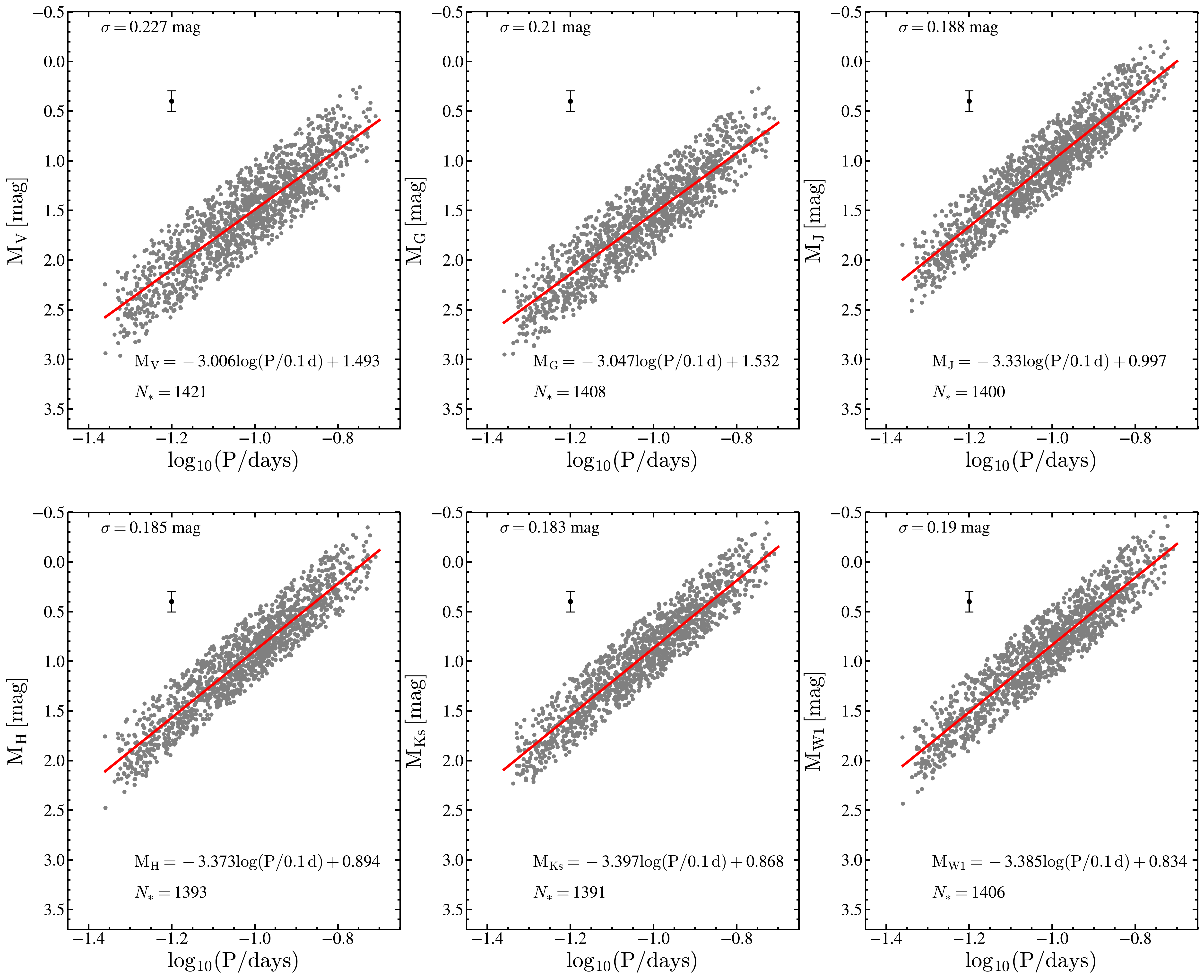}
    \caption{The PLRs for the fundamental mode $\delta$ Scuti stars in the $V$, Gaia DR2 $G$, $J$, $H$, $K_s$ and $W_1$ bands. The average uncertainty in $M_\lambda$ is shown in black. The fitted PLR is shown in red.}
    \label{fig:fig8}
\end{figure*}

\begin{figure*}
	\includegraphics[width=\textwidth]{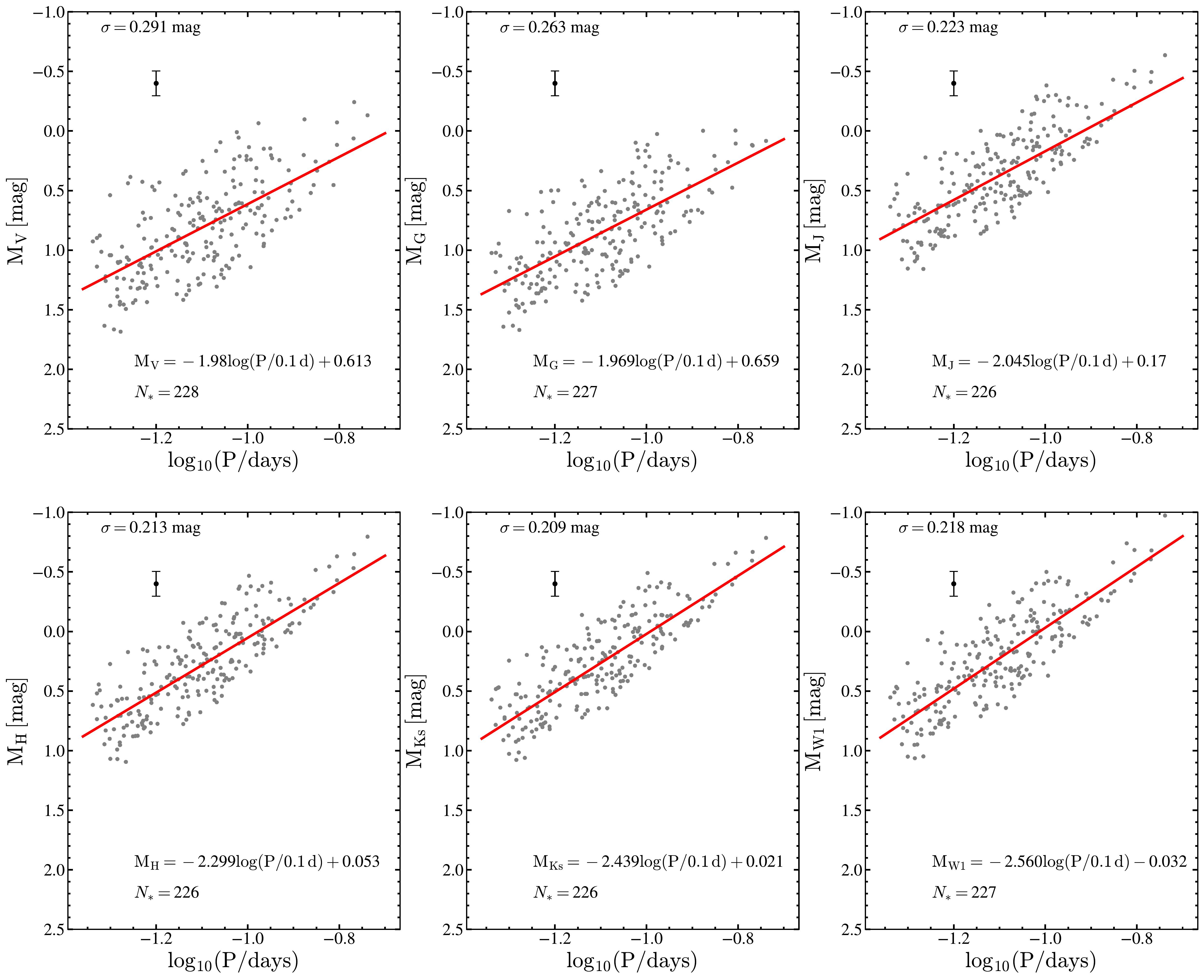}
    \caption{The PLRs for the overtone $\delta$ Scuti stars in the $V$, Gaia DR2 $G$, $J$, $H$, $K_s$ and $W_1$ bands. The average uncertainty in $M_\lambda$ is shown in black. The fitted PLR is shown in red.}
    \label{fig:fig9}
\end{figure*}

Next, after correcting for interstellar extinction with the SFD estimate, we perform PLR fits in the $V$, Gaia DR2 $G$, $J$, $H$, $K_s$ and $W_1$ bands for both the fundamental mode and overtone pulsators. The best-fit parameters, their uncertainties and the standard deviation from the PLR fit ($\sigma$) are listed in Table \ref{tab:fundfits} (fundamental) and Table \ref{tab:otfits} (overtone). We have illustrated these fits in Figure \ref{fig:fig8} (fundamental) and Figure \ref{fig:fig9} (overtone). The PLR derived in the $V$-band is consistent within uncertainties to those obtained by previous studies \citep{2019MNRAS.486.4348Z,2011AJ....142..110M}. The PLRs for the overtone pulsators have a larger scatter and uncertainty when compared to the corresponding fundamental mode PLRs. This is in part due to the relative rarity of the overtone pulsators when compared to the fundamental mode pulsators. In addition, the overtone sample likely encompasses sources with different dominant overtones, so the PLR fit is to an average distribution, and not that of a particular overtone. As is typical of PLRs, the near-infrared PLRs have smaller uncertainties and scatter than the PLRs in the optical. This effect is more pronounced for the overtone sources than for the fundamental mode pulsators. 

\begin{figure}
	\includegraphics[width=0.5\textwidth]{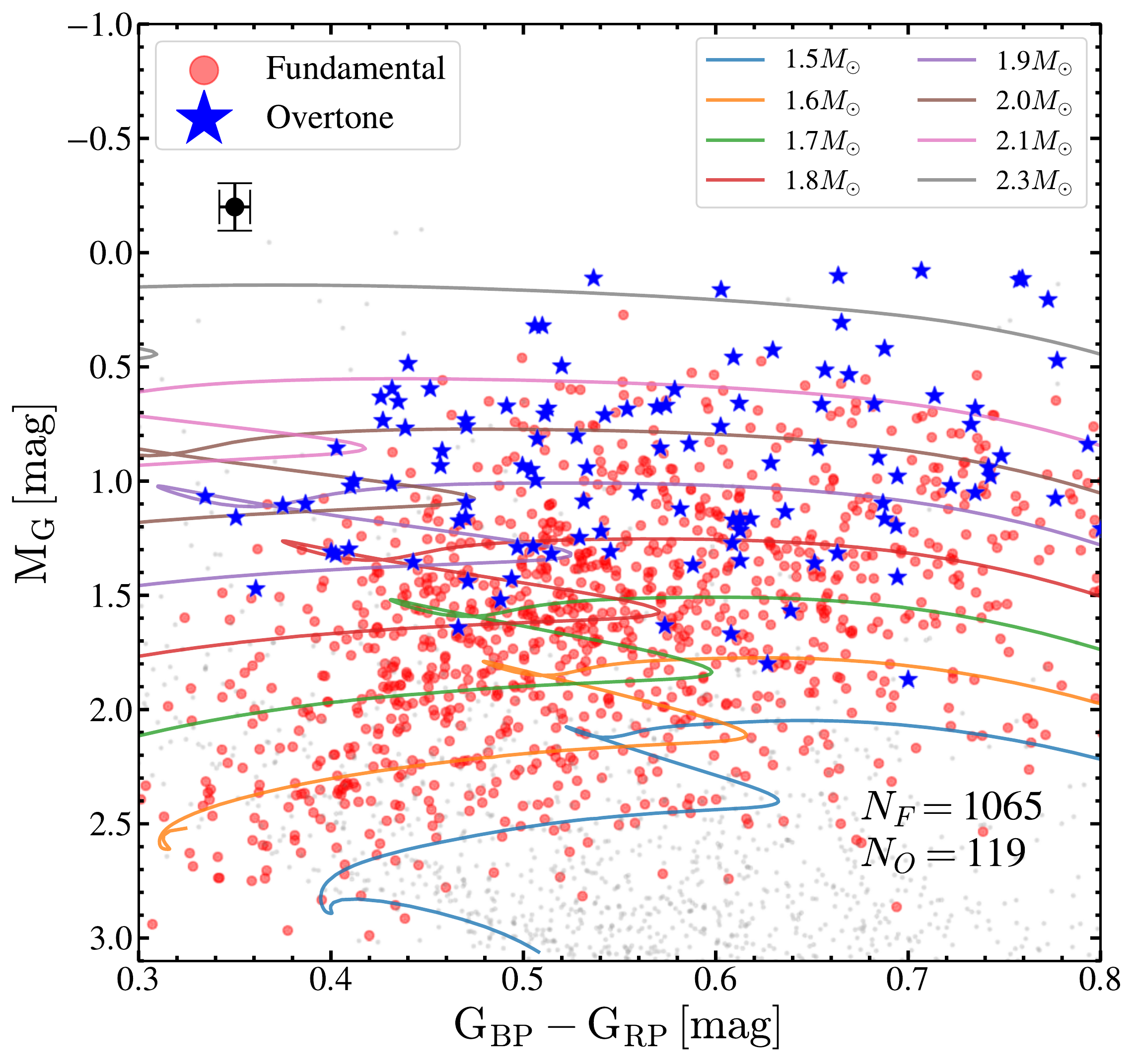}
    \caption{Gaia DR2 color-magnitude diagram for a sample of the fundamental mode and overtone pulsators with $A_V<0.5$ mag and parallaxes better than $10\%$. A sample of nearby sources with good parallaxes and photometry is shown in gray. Solar metallicity stellar evolution tracks from MIST \citep{2016ApJ...823..102C,2016ApJS..222....8D} are shown for comparison. The average uncertainties are shown in black.}
    \label{fig:fig10}
\end{figure}

Figure \ref{fig:fig10} shows the positions of a sample of the fundamental mode and overtone pulsators with $A_V<1$ mag and parallaxes better than $10\%$ in a Gaia DR2 color-magnitude diagram (CMD) after correcting for interstellar extinction. A sample of nearby sources with good parallaxes and photometry is shown in the background. To compare the observational data with theoretical models, we also show the MESA Isochrones and Stellar Tracks (MIST) stellar evolution models \citep{2016ApJ...823..102C,2016ApJS..222....8D} for stars with masses of 1.5 $M_{\odot}$, 1.6$M_{\odot}$, 1.7 $M_{\odot}$, 1.8 $M_{\odot}$, 1.9 $M_{\odot}$, 2.0 $M_{\odot}$, 2.1 $M_{\odot}$, and 2.3 $M_{\odot}$ starting from the ZAMS. We assumed Solar metallicity and no extinction for these models. The overtone pulsators largely have $M_G>2$ mag and appear to be consistent with the stellar evolution tracks with masses $1.8M_{\odot}<M<2.1 M_{\odot}$. In contrast, the fundamental mode pulsators appear to span a wider range of masses and luminosities that overlap with the overtone pulsators. 

We investigate how the fundamental mode $W_{JK}$ PLR varies with pulsational amplitude and the Gaia DR2 distances. We studied the variation of the PLR fits in amplitude bins with $\rm A_{RFR}<0.15$ mag, 0.15 mag $\rm <A_{RFR}<0.30$ mag, and $\rm A_{RFR}>0.30$ mag and found that the $W_{JK}$ PLR fits remained consistent with the overall fit (Table \ref{tab:fundfits}). We also investigated the variation of the PLR fits in distance bins with $\rm D<1$ kpc, 1 kpc $<\rm D<2$ kpc, and $\rm D>2$ kpc. The $W_{JK}$ PLR fits remain consistent with the overall fit in the bins $\rm D<1$ kpc and 1 kpc $<\rm D<2$ kpc, but the PLR fit for the bin $\rm D>2$ kpc differs slightly from the overall fit. This is likely due to the decreasing reliability of Gaia DR2 distances at scales larger than ${\sim}1-2$ kpc which introduces additional vertical scatter in the PLR.

In addition, \citet{2019MNRAS.485.2380M} noted that the ${\sim}29\, \mu as$ global zero-point parallax offset implemented in the distance estimates by \citet{2018AJ....156...58B} resulted in underestimated luminosities for A-type stars. To investigate the effects of this parallax offset on the PLRs, we recalculated the PLRs by selecting sources with \verb"parallax"/\verb"parallax_error" $>20$ where the effect of a parallax offset should be minimal. For these sources, this offset corresponds to roughly ${\sim}1\times$ the parallax uncertainties. We find that the parameters of the PLR fits change by $<2\%$. Since the PLR fits are not dramatically different, the impact of the zero-point parallax offset on the PLRs should be negligible.

\newpage

\section{The spectroscopic sub-sample}

\begin{figure*}
	\includegraphics[width=\textwidth]{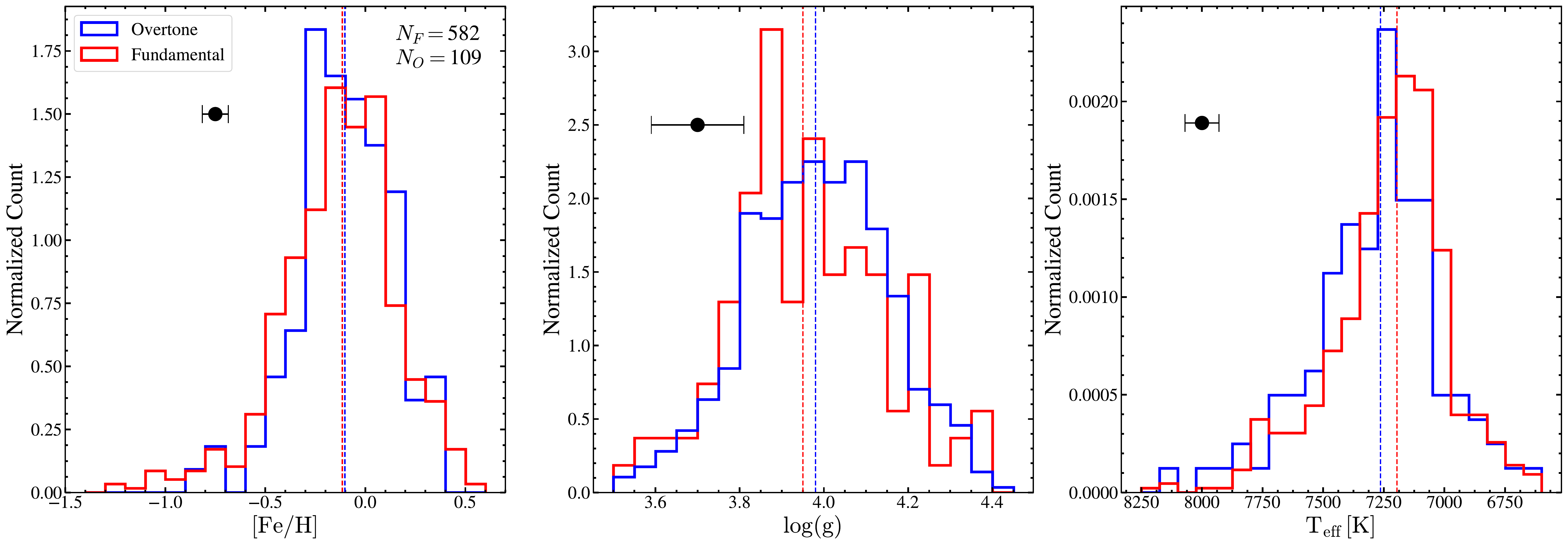}
    \caption{Distributions of the fundamental mode (red) and overtone (blue) $\delta$ Scuti stars in $\rm [Fe/H]$, $\rm \log (g)$, and $\rm T_{eff}$. The median value for each parameter is illustrated with a dashed line. The average uncertainty for each parameter is shown in black.}
    \label{fig:fig11}
\end{figure*}

We found 972 cross-matches to LAMOST, GALAH or RAVE. Figure \ref{fig:fig11} shows the distributions in effective temperature $\rm T_{eff}$, surface gravity $\rm \log (g)$ and metallicity $\rm [Fe/H]$ for both the fundamental mode and overtone $\delta$ Scuti. Given the relatively small number of sources with spectroscopic data, we relax the cutoffs implemented in $\S3.1$ and consider sources with $\rm Prob>0.9$, and \verb"parallax"/\verb"parallax_error" $>5$. The distributions of the fundamental ($N=582$) and overtone ($N=109$) $\delta$ Scuti stars in $\rm \log (g)$ and $\rm [Fe/H]$ are consistent with each other. The median $\rm T_{eff}$ for the overtone sources is lower than the median $\rm T_{eff}$ for the fundamental mode sources by ${\sim}60$K, but this difference is small given the typical uncertainties in the $\rm T_{eff}$ measurements. SX Phoenicis (SX Phe) stars are considered the type II analogues of the $\delta$ Scuti stars with typical metallicities $\rm [Fe/H]<-1$ \citep{1995AJ....109.1751M,2012MNRAS.419..342C}. From the cross-matches, 15 had $\rm [Fe/H]<-1$, thus we estimate the contamination rate of the SX Phe variables as ${\sim}1.5\%$.  On average, the $\delta$ Scuti stars with spectroscopic data have median $\rm [Fe/H]{\sim}-0.11$, $\rm \log (g){\sim}4.0$, and $\rm T_{eff}{\sim}7250$ K.

\begin{figure*}
	\includegraphics[width=\textwidth]{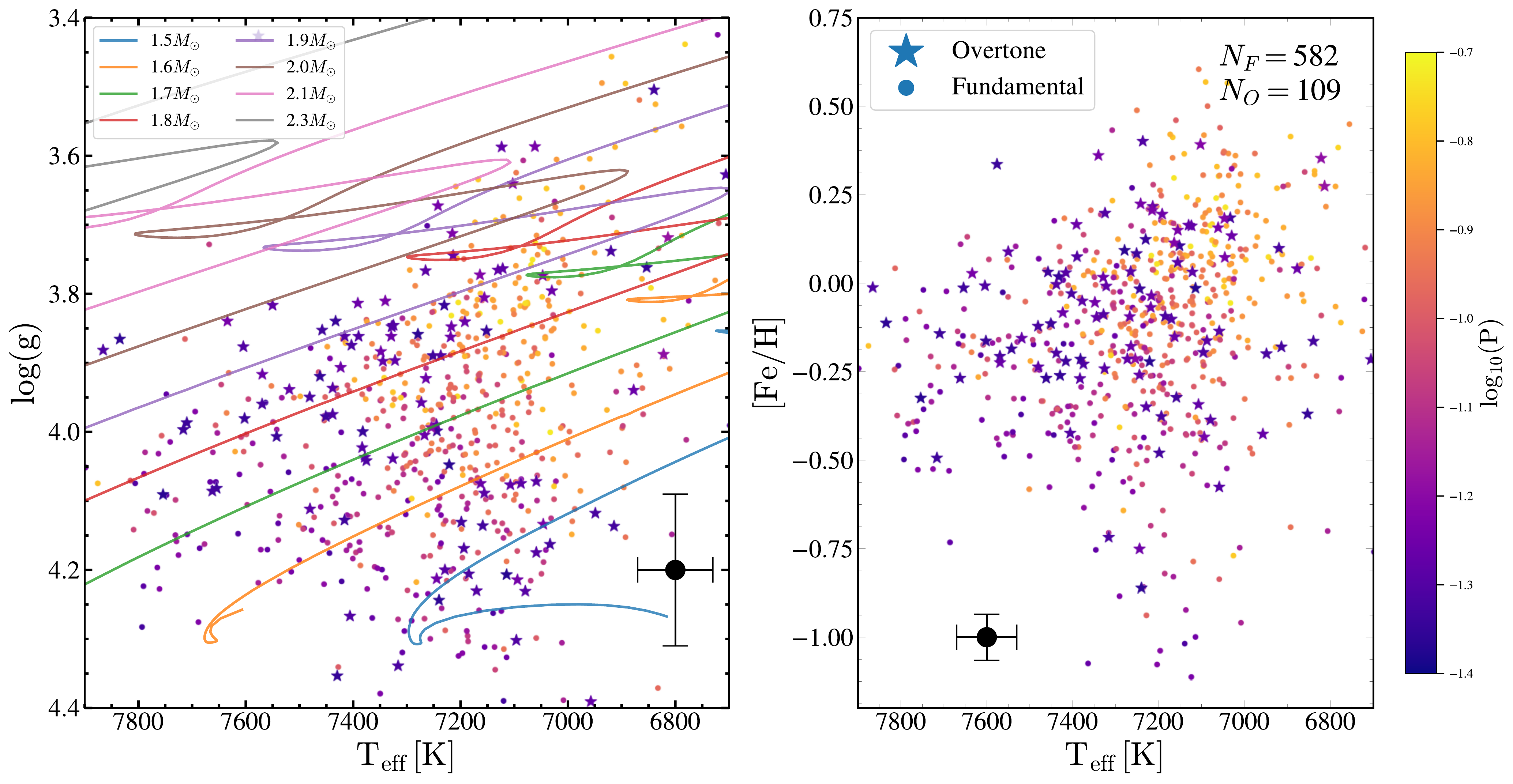}
    \caption{Distributions of the fundamental mode and overtone $\delta$ Scuti stars in $\rm \log (g)$ vs. $\rm T_{eff}$ (left) and $\rm T_{eff}$ vs. $\rm [Fe/H]$ (right).  The points are colored by the pulsational period. Solar metallicity stellar evolution tracks from MIST \citep{2016ApJ...823..102C,2016ApJS..222....8D} are shown for comparison. The average uncertainties are shown in black. }
    \label{fig:fig12}
\end{figure*}

\begin{figure*}
	\includegraphics[width=\textwidth]{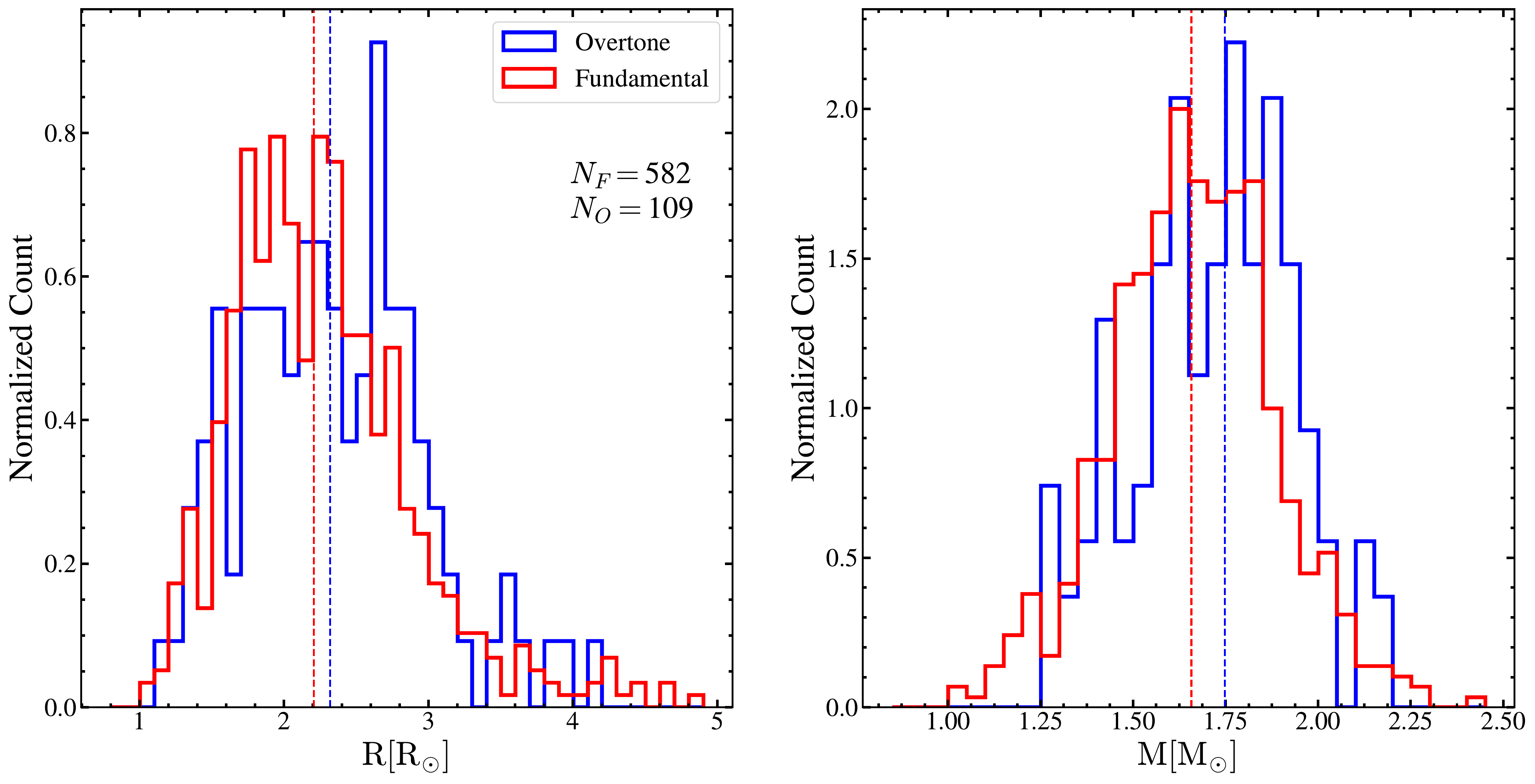}
    \caption{Distributions of radii (left) and masses (right) of the fundamental mode (red) and overtone (blue) $\delta$ Scuti stars. The median value for each parameter is illustrated with a dashed line.}
    \label{fig:fig13}
\end{figure*}

\begin{figure}
	\includegraphics[width=0.5\textwidth]{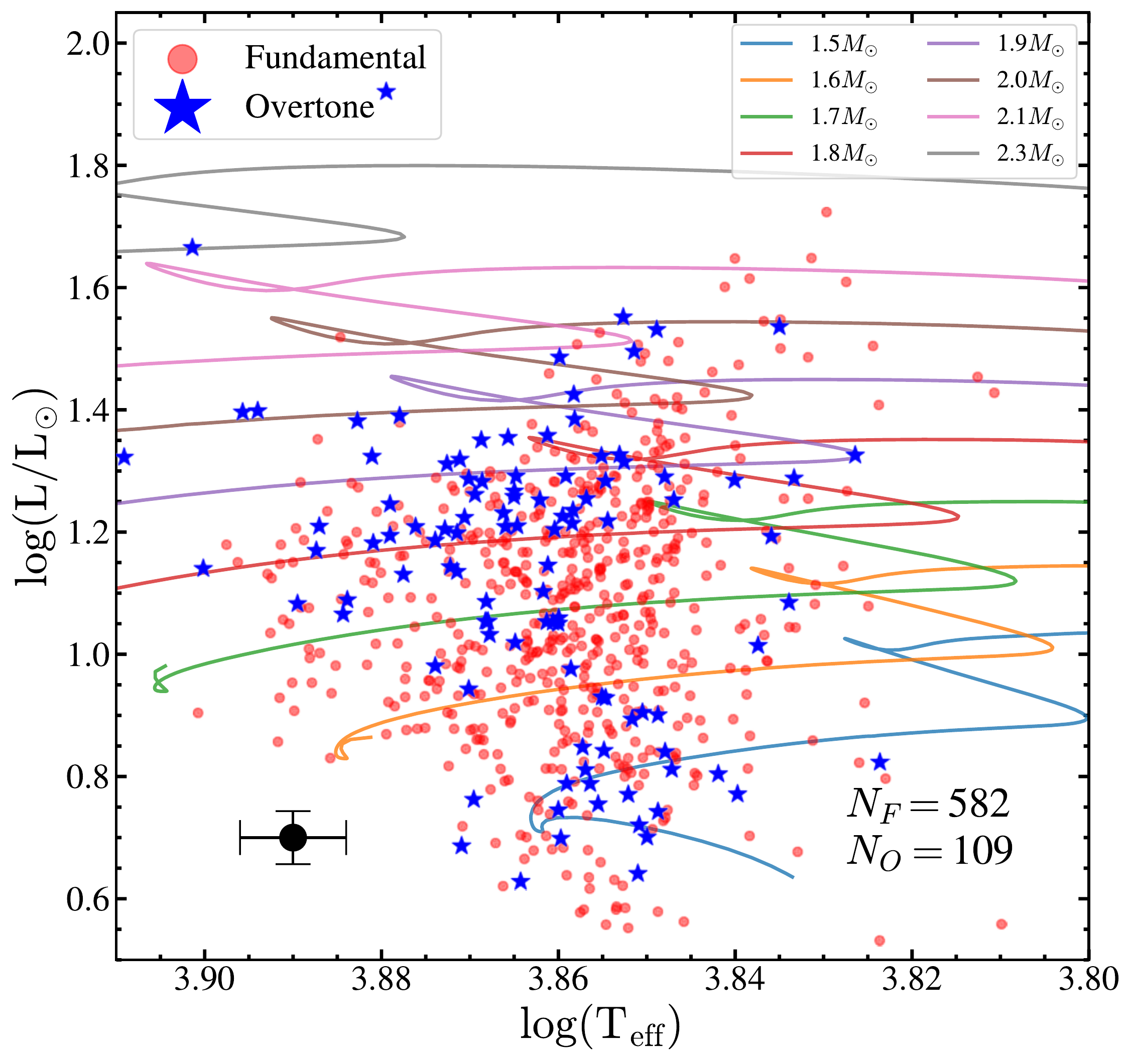}
    \caption{Hertzsprung-Russell diagram for the spectroscopic sub-sample of the fundamental mode and overtone pulsators. Solar metallicity stellar evolution tracks from MIST \citep{2016ApJ...823..102C,2016ApJS..222....8D} are shown for comparison.}
    \label{fig:fig14}
\end{figure}

We further look at the correlations between these spectroscopic parameters in Figure \ref{fig:fig12}. We compare the observed data with the MIST stellar evolution models described in $\S3.1$. These models agree very well with the observed values of $\rm T_{eff}$ and $\rm \log (g)$. Most of our sources have $\rm T_{eff}$ and $\rm \log (g)$ consistent with stars having masses $M<2.1M_{\odot}$. We use the calibrations from \citet{2010A&ARv..18...67T} to derive the masses and radii of the $\delta$ Scuti stars in the spectroscopic sub-sample using $\rm T_{eff}$, $\rm \log (g)$ and $\rm [Fe/H]$. The luminosities are then derived using the radius and $\rm T_{eff}$. Figure \ref{fig:fig13} shows the distribution of masses and radii of these sources. The median radii of the fundamental mode and overtone pulsators are similar, with $\rm \langle R_{fundamental}\rangle{\sim}2.2 R_\odot$ and $\rm \langle R_{overtone}\rangle{\sim}2.3 R_\odot$. The distribution of masses of the overtone sources seem to suggest an excess of sources at higher mass bins ($M>1.75M_{\odot}$) than the fundamental mode pulsators, which is consistent with the position of these stars in the Gaia DR2 CMD. The median masses of the fundamental mode and overtone pulsators are slightly different, with $\rm \langle M_{fundamental}\rangle{\sim}1.66 M_\odot$ and $\rm \langle M_{overtone}\rangle{\sim}1.75 M_\odot$. However, we note that this difference is small given the uncertainties of the masses derived using the \citet{2010A&ARv..18...67T} calibration. Figure \ref{fig:fig14} shows the Hertzsprung-Russell diagram for these sources. This data is also consistent with most of these sources having masses $M<2.1M_{\odot}$.

\begin{figure*}
	\includegraphics[width=\textwidth]{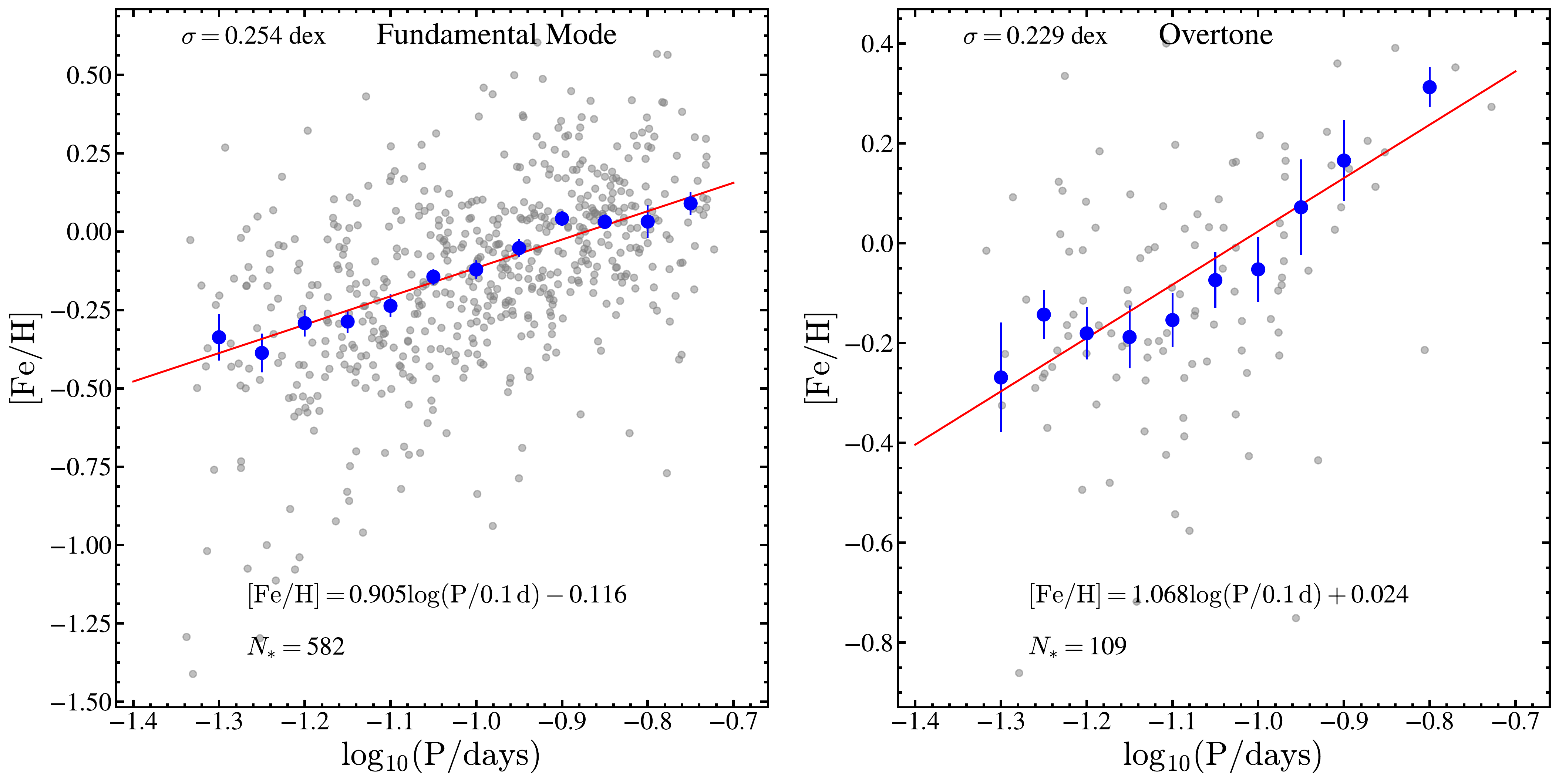}
    \caption{Distribution of median metallicity with $\log _{10} (\rm P/days)$ for the fundamental mode (left) and overtone (right) pulsators. The individual fundamental mode and overtone pulsators used in the binning are shown in gray. A linear fit to the binned data is shown in red.}
    \label{fig:fig15}
\end{figure*}

In Figure \ref{fig:fig12}, there is a strong correlation between period and composition. Fundamental mode $\delta$ Scuti stars with near-Solar metallicities have longer periods than lower metallicity sources. Figure \ref{fig:fig15} shows the dependence of the median metallicity on $\log _{10} (\rm P/days)$ for both the fundamental mode and overtone pulsators. Linear fits give \begin{equation}
    \rm [Fe/H]_{F}=\rm 0.908(\pm0.062)\log_{10}(P/0.1\,d)-0.117(\pm0.011),
	\label{eq:fehfund}
\end{equation} for the fundamental mode and\begin{equation}
    \rm [Fe/H]_{O}=\rm 1.068(\pm0.134)\log_{10}(P/0.1\,d)+0.024(\pm0.022),
	\label{eq:fehot} 
\end{equation} for the overtone pulsators. These relationships both have large scatter ($\sigma{\sim}0.25$ dex). The $\rm [Fe/H]$ - $\log _{10} (\rm P/days)$ slopes of the two fits are consistent given the uncertainties, but the zero points differ as expected. At any given period, the overtone pulsators have a higher metallicity than fundamental mode pulsators of the same period.

\begin{figure*}
	\includegraphics[width=\textwidth]{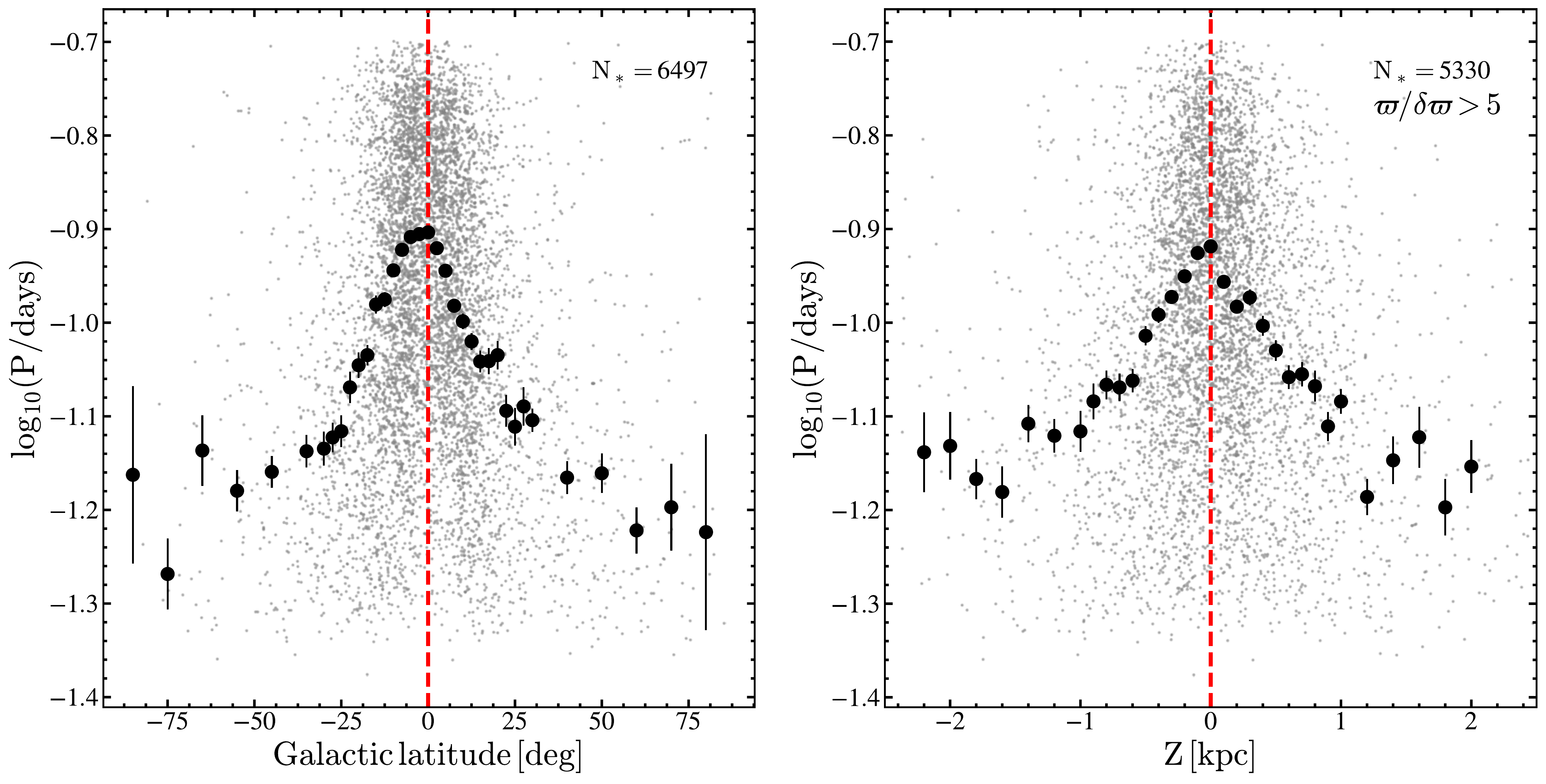}
    \caption{Distribution of median periods with Galactic latitude (left) in bins of $\rm 2.5\, deg$ ($\rm -30\, deg\leq b\leq 30\, deg$) and $\rm 10\, deg$ ($\rm |b|> 30\, deg$), and the distance from the Galactic mid-plane (right) in bins of 100 pc ($\rm -1\,kpc\leq Z\leq 1\,kpc$) and 200 pc ($\rm |Z|> 1\,kpc$). The individual $\delta$ Scuti stars used in the binning are shown in gray. }
    \label{fig:fig16}
\end{figure*}

\begin{figure}
	\includegraphics[width=0.5\textwidth]{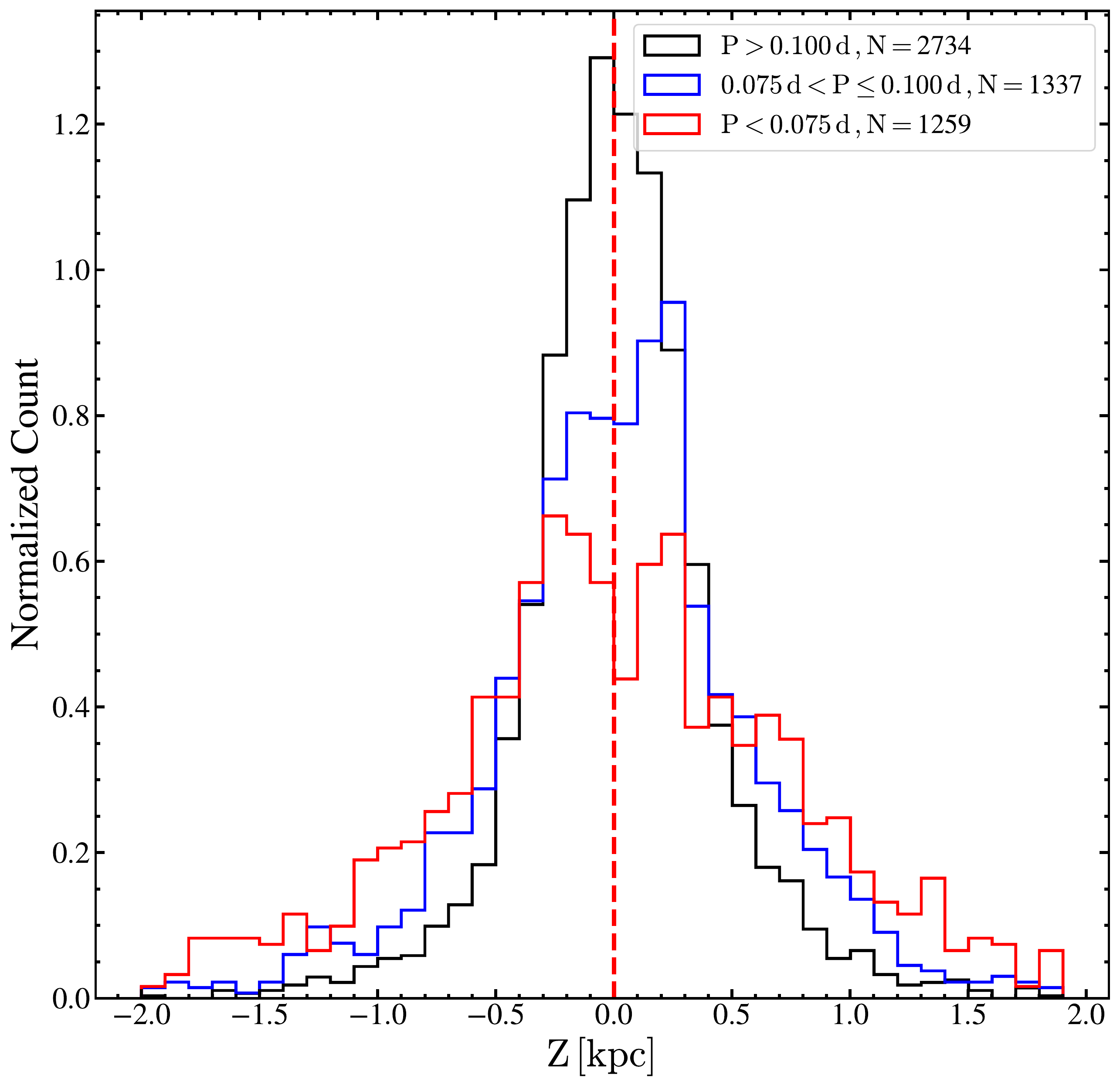}
    \caption{Distribution of the distances from the Galactic mid-plane in bins of 100 pc ($\rm -1\,kpc\leq Z\leq 1\,kpc$) and 200 pc ($\rm |Z|> 1\,kpc$) for $\delta$ Scuti stars with periods $\rm P<0.075\,d$ (red), $\rm 0.075\,d<P\leq 0.100\,d$ (blue), and $\rm P>0.100\,d$ (black).}
    \label{fig:fig17}
\end{figure}

The median period varies significantly with metallicity, from $\log _{10} (\rm P/days){\sim}-1.1$ for sources with $\rm [Fe/H]<-0.3$ to $\log _{10} (\rm P/days){\sim}-0.9$ for sources with super-solar metallicities $\rm [Fe/H]>0$. This is reflected in the positions of the $\delta$ Scuti stars in the Galaxy (Figure \ref{fig:fig2}). Sources closer to the Galactic disk, which should have higher metallicities \citep{2009A&A...504...81P}, tend to have longer periods than sources further away from the disk (Figure \ref{fig:fig16}). Thus, the $\delta$ Scuti period gradient in Galactic latitude is likely due to the vertical metallicity gradient of the Galactic disk. To characterize this effect further, we calculate the cylindrical Galactocentric coordinates ($\rm R$ and $\rm Z$) for these sources using the Gaia DR2 distances (parallaxes better than $20\%$) and the \verb"astropy" coordinate transformations. Figure \ref{fig:fig16} shows the median periods as a function of the distance from the Galactic mid-plane ($\rm Z$) in bins of 100 pc ($\rm -1\,kpc\leq Z\leq 1\,kpc$) and 200 pc ($\rm |Z|> 1\,kpc$). It is evident that towards the Galactic disk, the median period increases, whereas towards the halo, the median period drops sharply. Figure \ref{fig:fig17} shows the distribution of distances from the Galactic mid-plane for $\delta$ Scuti stars with periods $\rm P<0.075\,d$, $\rm 0.075\,d<P\leq 0.100\,d$, and $\rm P>0.100\,d$. We find that there is an excess of sources with short periods ($\rm P<0.100\,d$) away from the Galactic disk ($\rm |Z|> 0.5\,kpc$) when compared to the sources with longer periods. This effect is more pronounced at even shorter periods as the sources with $\rm P<0.075\,d$ have a broader distribution in $\rm Z$ than the sources with $\rm 0.075\,d<P\leq 0.100\,d$. Stars with $\rm P>0.100\,d$ are predominantly located close to the Galactic disk ($\rm |Z|<0.5\,kpc$).
    
This could potentially be an observational bias related to the decreasing sensitivity to lower luminosity (i.e., shorter period) $\delta$ Scuti stars towards the Galactic disk due to extinction. To investigate this further, we examined the distributions of $W_{JK}$ magnitudes along with the distributions of the median period with distance from the Galactic mid-plane in extinction bins of $\rm A_V<0.5\,mag$, $\rm 0.5\,mag<A_V\leq 1\,mag$ and $\rm A_V>1\,mag$. Here, we used variable binning in $\rm Z$ as the sources with high extinctions ($\rm A_V>0.5\,mag$) are largely located towards the disk ($\rm |Z|<1\,kpc$), whereas the sources with small extinctions are located away from the disk. The same trend observed in Figure \ref{fig:fig16} is reproduced in all the extinction bins.

\begin{figure*}
	\includegraphics[width=\textwidth]{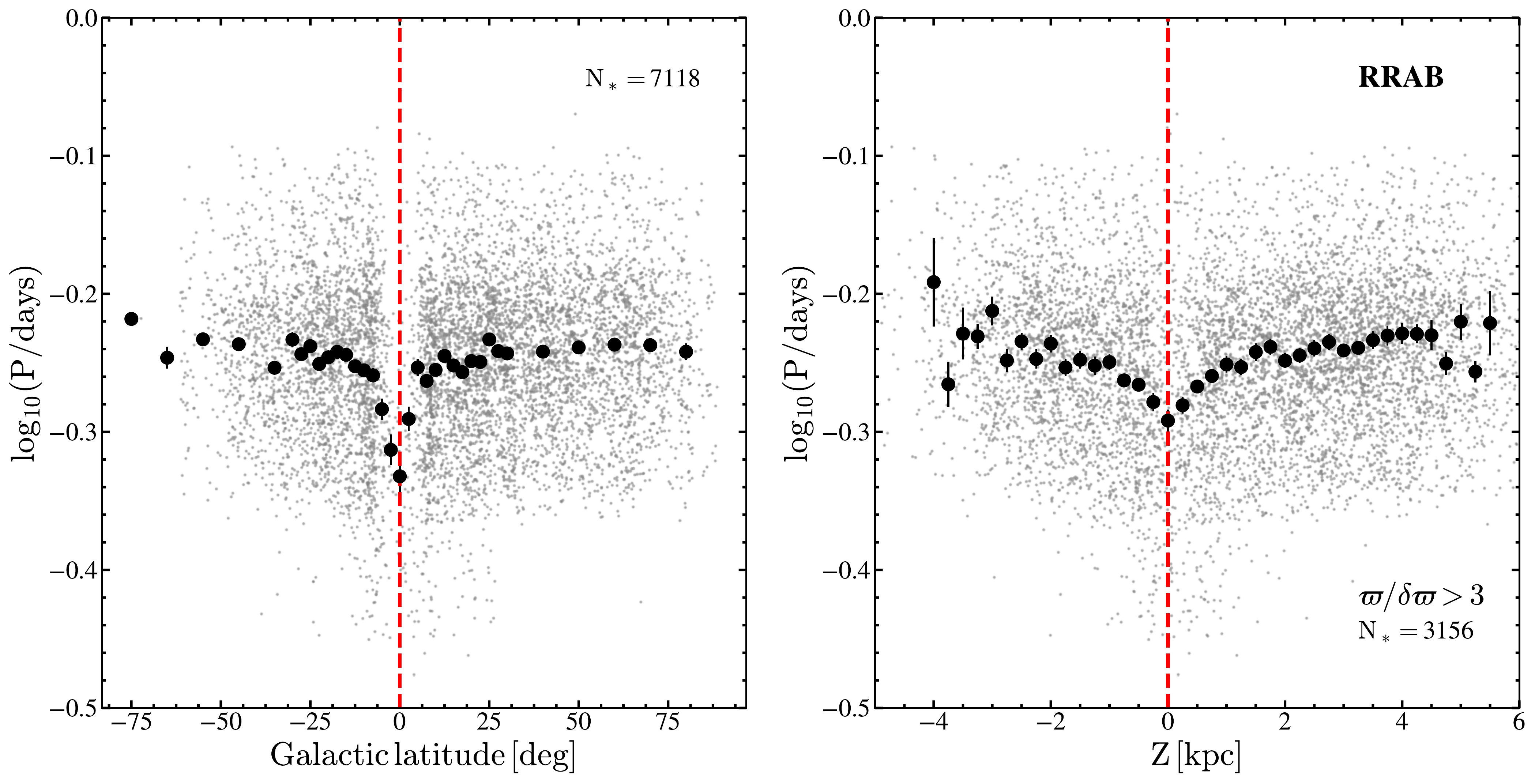}
    \caption{Distribution of the median periods of fundamental mode RR Lyrae stars (RRab) with Galactic latitude (left) in bins of $\rm 2.5\, deg$ ($\rm -30\, deg\leq b\leq 30\, deg$) and $\rm 10\, deg$ ($\rm |b|> 30\, deg$), and the distance from the Galactic mid-plane (right) in bins of 250 pc ($\rm -3.5\,kpc\leq Z\leq 4\,kpc$). The individual RRab stars used in the binning are shown in gray. As expected, this is the opposite trend seen in $\delta$ Scuti stars (Figure \ref{fig:fig16}).}
    \label{fig:fig18}
\end{figure*}

As a check on the vertical metallicity gradient explaining the vertical trend in period, we can examine the trend for RR Lyrae stars. Metal rich RRab stars have shorter periods \citep{1993AJ....106..687S,2018SSRv..214..113B}, so a vertical metallicity gradient should lead to a reversed trend in period with $\rm Z$ from the $\delta$ Scuti. We selected a sample of fundamental mode RR Lyrae (RRAB) stars from the ASAS-SN catalogue with \verb"parallax"/\verb"parallax_error" $>3$, and we see exactly this reversed correlation in Figure \ref{fig:fig18}. We will explore such trends further in Jayasinghe et al. (2020, in prep).

\begin{figure*}
	\includegraphics[width=\textwidth]{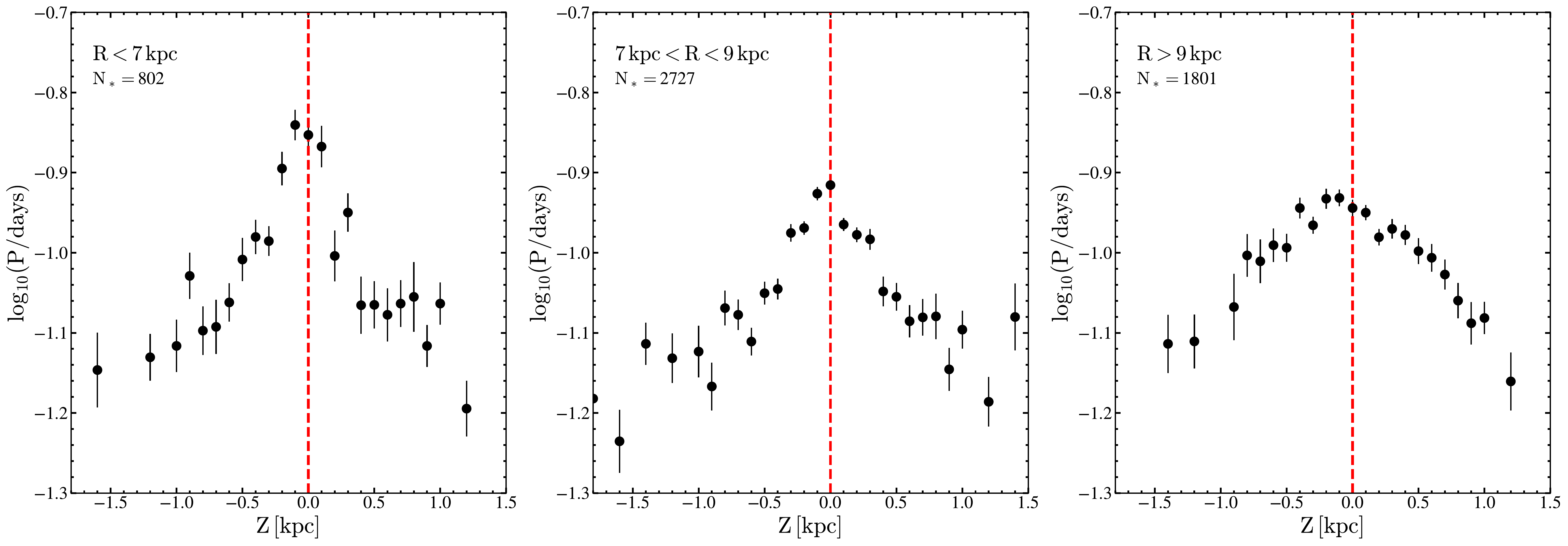}
    \caption{Distribution of median periods with distances from the Galactic mid-plane in three bins of Galactocentric radius with $\rm R<7\, kpc$ (left), $\rm 7\,kpc<R<9\, kpc$ (center) and $\rm R>9\, kpc$ (right).}
    \label{fig:fig19}
\end{figure*}

Figure \ref{fig:fig19} shows the median periods as a function of the distance from the Galactic mid-plane ($\rm Z$) in bins of 100 pc ($\rm -1\,kpc\leq Z\leq 1\,kpc$) and 200 pc ($\rm |Z|> 1\,kpc$) for sources with Galactocentric radii $\rm R<7\, kpc$, $\rm 7\,kpc<R<9\, kpc$ and $\rm R>9\, kpc$. We see that at fixed $\rm |Z|$, stars at lower Galactocentric radii have longer periods, consistent with the expected radial gradient of metallicity \citep{2009A&A...504...81P}. The trends are harder to illustrate, however, because the stars span a limited fractional range in R and there are strong correlations between R and the average height above the plane. The median period at $\rm Z{\sim}0$ increases from $\log _{10} (\rm P/days){\sim}-0.94$ in the bin $\rm R>9\, kpc$ to $\log _{10} (\rm P/days){\sim}-0.85$ in the bin $\rm R<7\, kpc$. We also find that the widths of these distributions broaden with increasing distance from the Galactic center, which is consistent with our finding that longer period $\delta$ Scuti stars are more likely to be located towards regions with higher metallicity.

With measurements of $\rm T_{eff}$ and $\rm \log (g)$, we are also able to calculate the pulsational constant $$Q=P\sqrt{\frac{\bar{\rho}}{\bar{\rho}_{\odot}}},$$ where $P$ is the period in days and $\bar{\rho}/\bar{\rho}_{\odot}$ is the mean stellar density normalized by the mean Solar density \citep{2017ampm.book.....B}. This can be rewritten as $$\log(Q)=\log(P) + 0.5\log(g)+0.1 M_{bol}+\log(T_{\rm eff})-6.454,$$ where $M_{bol}=M_V-A_V+BC$ is the absolute bolometric magnitude of the source \citep{2017ampm.book.....B}. We derived the bolometric correction for these sources using the \verb"isochrones" package \citep{2015ascl.soft03010M} which utilizes the MIST bolometric correction grids \citep{2016ApJ...823..102C,2016ApJS..222....8D}.

\begin{figure*}
	\includegraphics[width=\textwidth]{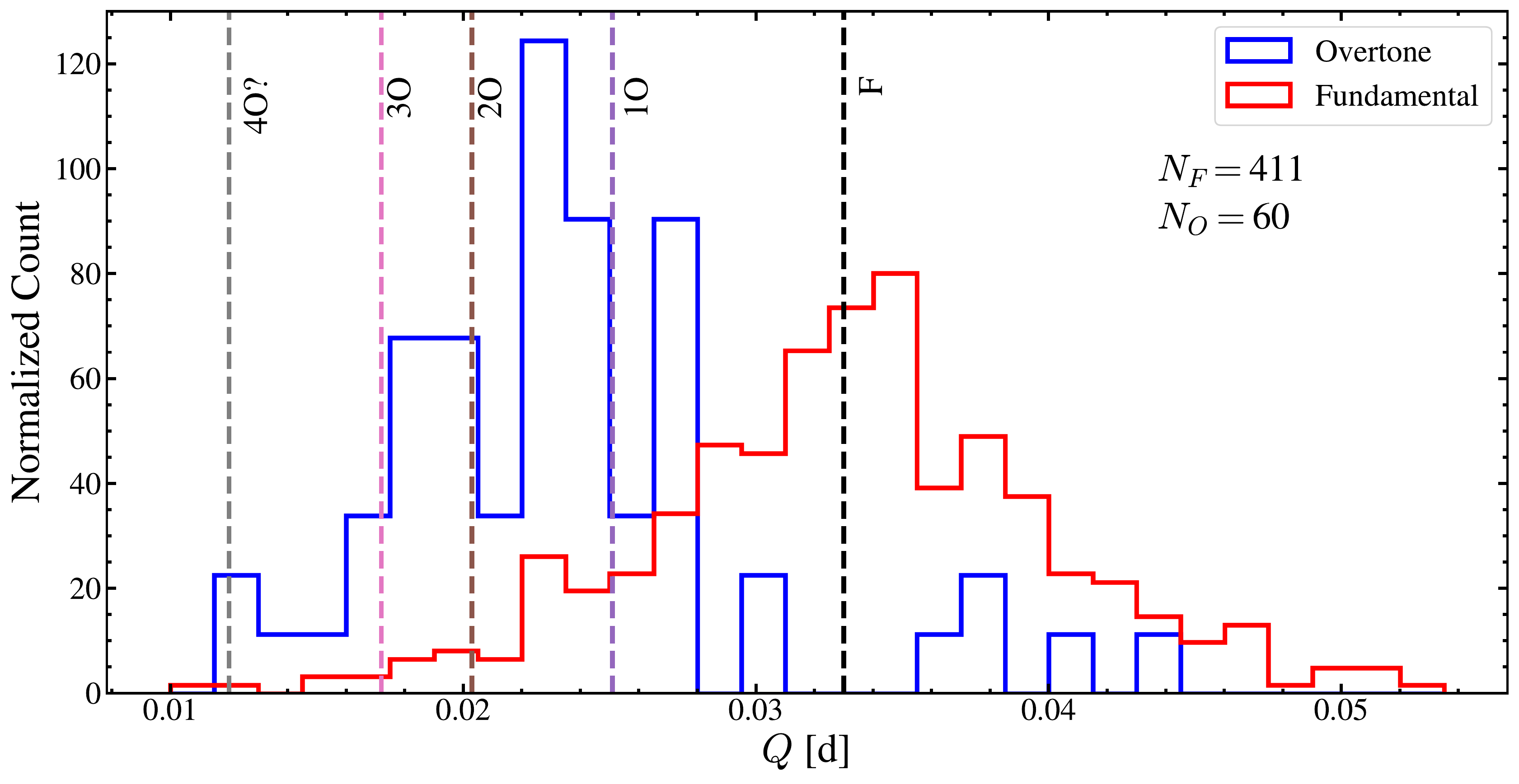}
    \caption{Distribution of the pulsational constants ($Q$) for the sources in the spectroscopic sub-sample. The expected $Q$ values for the fundamental mode and different overtone modes from \citet{1997ESASP.402..367N} are also shown for reference.}
    \label{fig:fig20}
\end{figure*}

Figure \ref{fig:fig20} shows the distribution of $Q$ for both the fundamental mode and overtone pulsators in the spectroscopic sub-sample. The expected values of  $Q$ for the fundamental and overtone modes estimated by \citet{1997ESASP.402..367N} are also shown for reference. The calculation of $Q$ is strongly dependent on the stellar parameters \citep{1990A&A...231...56B,2017ampm.book.....B}, and the fractional uncertainties in $Q$ can be rather large ($\gtrsim 15\%$). Nevertheless, there is a clear distinction between the overtone pulsators and the fundamental mode pulsators. Most of the fundamental mode pulsators have $Q$ values close to the expected value for the radial fundamental mode ($Q_F{\sim}0.033$ d), but there is also evidence that some of these sources pulsate in the first overtone ($Q_{\rm 1O}{\sim}0.0255$ d). This suggests that some fraction of the sources classified as fundamental mode pulsators actually have a dominant period corresponding to the first overtone. This was already hinted at in Figure \ref{fig:fig7}. This uncertainty in the dominant pulsational mode between fundamental mode and first overtone pulsators adds to the scatter in the derived PLRs for the fundamental mode pulsators. This problem is difficult to solve with just the ASAS-SN data, as mode identification for a large ensemble of $\delta$ Scuti stars, most of which do not have spectroscopic determinations of $\rm T_{eff}$ and $\rm \log (g)$ or high cadence (\textit{Kepler/K2} or \textit{TESS}) data, is challenging. Overtone pulsators in the spectroscopic sub-sample form a distinct population in $Q$, with multiple peaks corresponding to the first overtone, second overtone ($Q_{\rm 2O}{\sim}0.0205$ d) and third overtone ($Q_{\rm 3O}{\sim}0.0174$ d) visible in the data, and an absence of sources with $Q$ close to the fundamental mode. This is also consistent with our observations in Figure \ref{fig:fig7}. The sample appears to be dominated by sources pulsating in the first overtone, followed by sources pulsating in the second and third overtones. We also note a peak centered at ($Q{\sim}0.012$ d) which could be the fourth overtone, but we were unable to find a prediction for this mode's $Q$ in the literature.

\section{$\delta$ Scuti stars in \textit{TESS}}

The Transiting Exoplanet Survey Satellite (\textit{TESS}; \citealt{2015JATIS...1a4003R}) is currently observing most of the sky with a baseline of at least 27 days and has finished observations in the Southern Hemisphere. \citet{2019arXiv190912018A} presented the first asteroseismic results for $\delta$ Scuti and $\gamma$ Dor stars observed in the \textit{TESS} mission, utilizing the 2-min cadence light curves in Sectors 1 and 2. As an experiment, we extracted \textit{TESS} light curves of 10 fundamental and 10 overtone stars in the Southern Hemisphere.

We analyze the \textit{TESS} data using an image subtraction pipeline optimized for use with the FFIs. This pipeline is derived from that used to process ASAS-SN data, and is based on the ISIS package \citep{1998ApJ...503..325A,2000A&AS..144..363A}. This technique has previously been applied successfully to \textit{TESS} observations of a wide range of astrophysical objects including supernovae \citep{2019MNRAS.487.2372V,2019arXiv190402171F}, tidal disruption events \citep{2019ApJ...883..111H}, young stellar objects \citep{2019arXiv190910636H}, and the most extreme heartbeat star yet discovered \citep{2019arXiv190100005J}. For each sector we constructed a reference image from the first 100 FFIs of good quality obtained during that sector, excluding those with sky background levels or PSF widths above average for the sector. FFIs associated with non-zero data quality flags were also excluded. The primary outcome of this procedure is the exclusion of FFIs obtained during spacecraft momentum dumps (quality flag value 32) and those obtained during epochs heavily impacted by stray Earthlight or Moonlight (quality flag value 2048).

While the image subtraction technique is well-suited to producing differential light curves, the large pixel scale of \textit{TESS} makes it difficult to obtain reliable measurements of the flux of a given source in the reference image. Instead of relying on this, we have estimated the reference flux using the \verb"ticgen" software package \citep{ticgen,2018AJ....156..102S}. The \textit{TESS}-band magnitude estimates from \verb"ticgen" were converted into fluxes using an instrumental zero point of 20.44 electrons per second in the FFIs, based on the values provided in the TESS Instrument Handbook \citep{tesshandbook}. Flux was then added to the raw differential light curves such that the median of each target's observations matched the estimated reference value. This allowed us to produce the normalized flux light curves shown in Figures~\ref{fig:fig23} and \ref{fig:fig24}. The light curves do not include epochs where the observations were compromised by scattered light artifacts from the Earth or Moon.

\begin{figure*}
	\includegraphics[width=\textwidth]{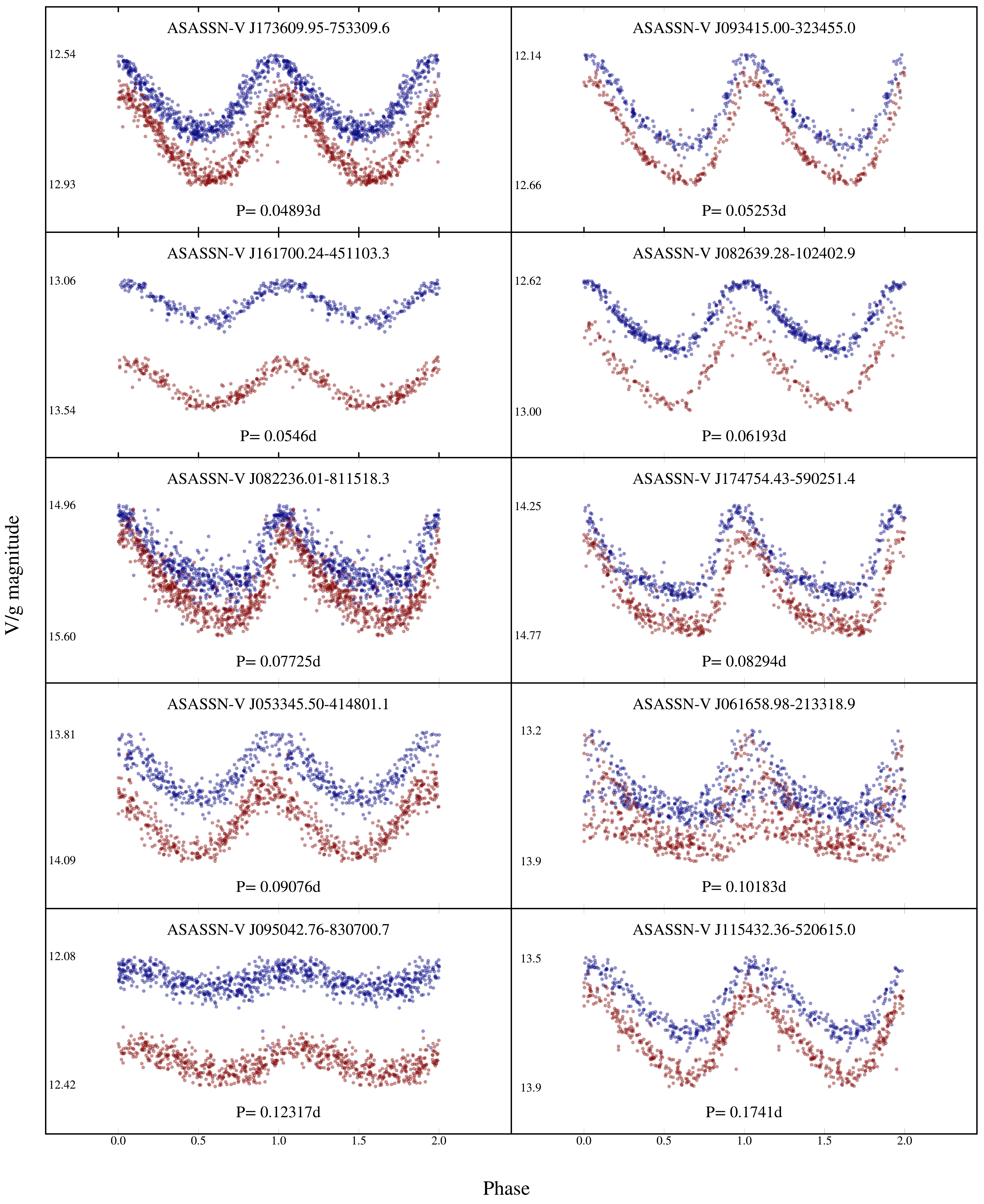}
    \caption{Phased ASAS-SN light curves for the 10 fundamental mode $\delta$ Scuti stars in the Southern Hemisphere with \textit{TESS} data. The light curves are scaled by their minimum and maximum V/g-band magnitudes. The blue points correspond to g-band data and the red points correspond to V-band data.}
    \label{fig:fig21}
\end{figure*}

\begin{figure*}
	\includegraphics[width=\textwidth]{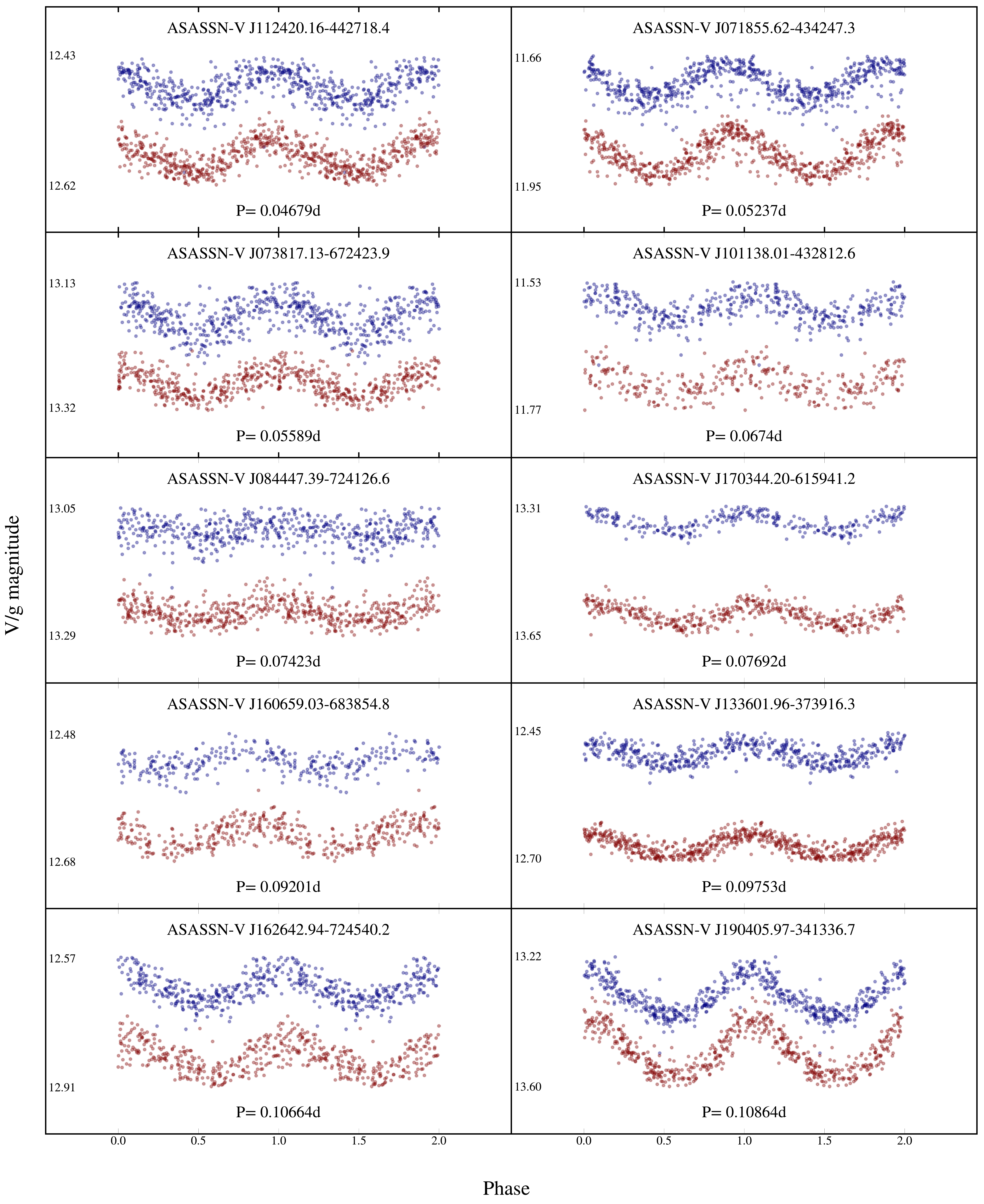}
    \caption{Phased ASAS-SN light curves for the 10 overtone $\delta$ Scuti stars in the Southern Hemisphere with \textit{TESS} data. The format is the same as Figure \ref{fig:fig21}. }
    \label{fig:fig22}
\end{figure*}

\begin{figure*}
	\includegraphics[width=\textwidth]{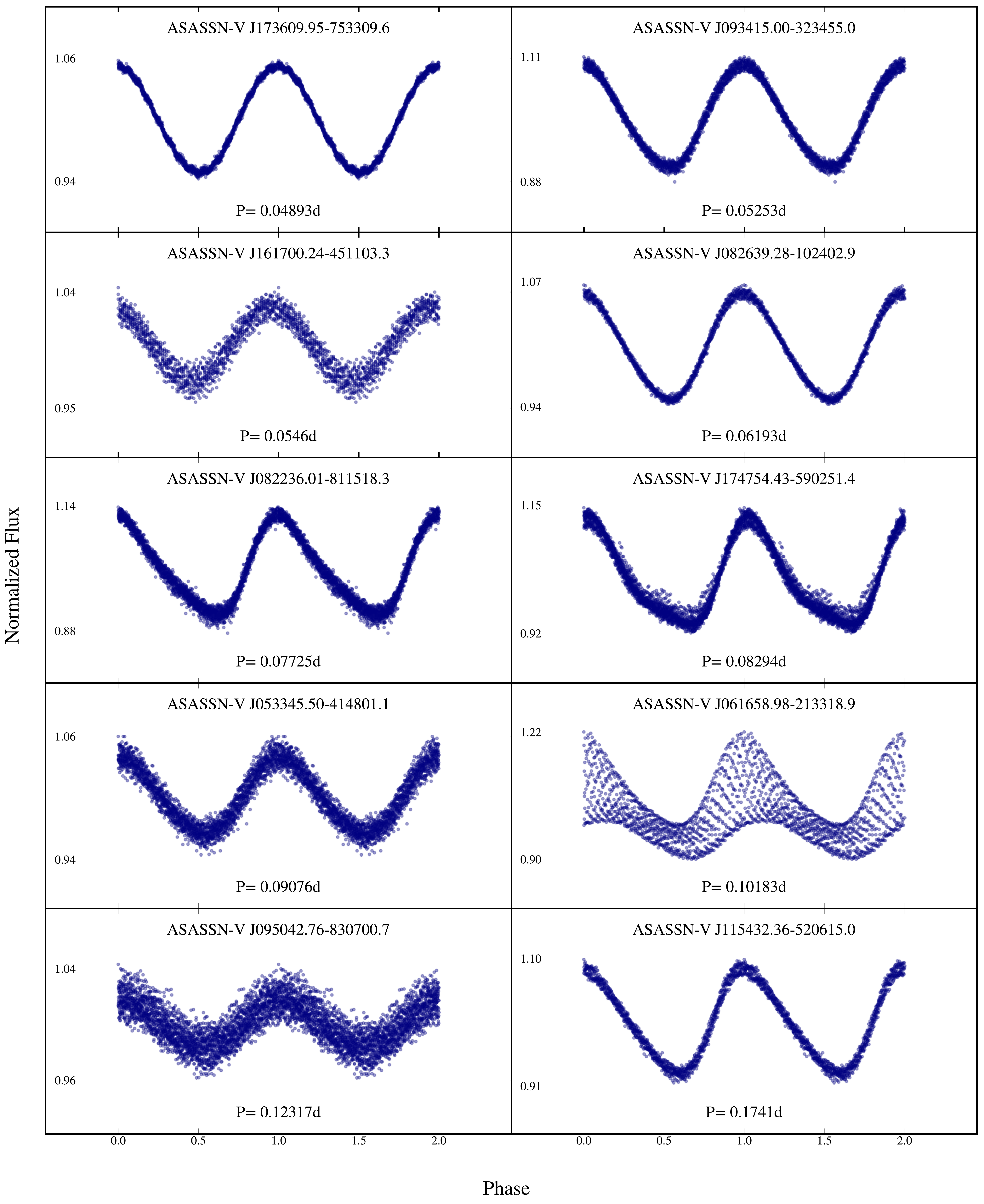}
    \caption{Phased \textit{TESS} light curves for the 10 fundamental mode $\delta$ Scuti stars in the Southern Hemisphere with \textit{TESS} data. The light curves are scaled by their minimum and maximum variations in flux. }
    \label{fig:fig23}
\end{figure*}

\begin{figure*}
	\includegraphics[width=\textwidth]{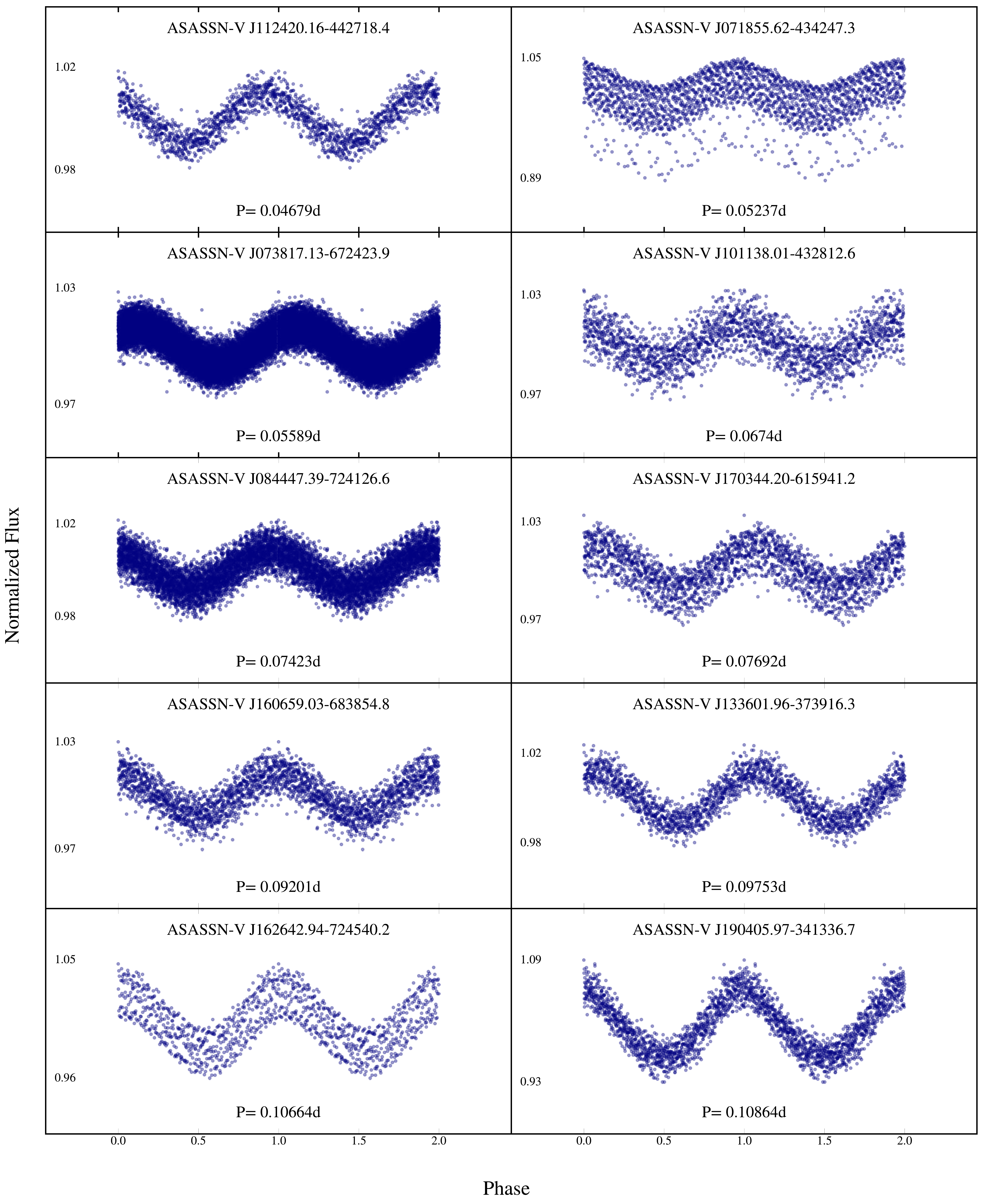}
    \caption{Phased \textit{TESS} light curves for the 10 overtone $\delta$ Scuti stars in the Southern Hemisphere with \textit{TESS} data. The format is the same as Figure \ref{fig:fig23}. }
    \label{fig:fig24}
\end{figure*}

These sample of 10 fundamental mode and 10 overtone $\delta$ Scuti stars with \textit{TESS} light curves are summarized in Table \ref{tab:srcinfo}. The phased ASAS-SN V- and g-band light curves are shown in Figure \ref{fig:fig21} (fundamental mode pulsators) and Figure \ref{fig:fig22} (overtone pulsators). The phased \textit{TESS }light curves for the same periods are shown in Figure \ref{fig:fig23} (fundamental mode pulsators) and Figure \ref{fig:fig24} (overtone pulsators).

The light curves of the fundamental mode pulsators are morphologically different from the light curves of the overtone pulsators. The overtone pulsators have more symmetric light curves but generally have multiple excited modes leading to more overall scatter. At this stage, we identified two sources with interesting light curves. ASASSN-V J061658.98$-$213318.9 has a light curve that suggests the existence of beat frequencies, and ASASSN-V J071855.62$-$434247.3 has evidence of eclipses.

\begin{table*}
	\centering
	\caption{Summary of the 10 fundamental mode and 10 overtone $\delta$ Scuti stars in the Southern Hemisphere with \textit{TESS} data}
	\label{tab:srcinfo}
\begin{tabular}{rrrrrrrrr}
		\hline
		 ASAS-SN ID & RAJ2000 & DEJ2000 & Period (d) & $\rm A_{RFR}$ (mag) & Type & $\Delta W_{JK}$ (mag) &  V (mag) & T (mag) \\
		\hline
J162642.94-724540.2 & 246.6789  & $-$72.76117 & 0.106638 & 0.13 & Overtone    & $-$0.68 & 12.59 & 12.20  \\
J084447.39-724126.6 & 131.19746 & $-$72.69072 & 0.074229 & 0.07 & Overtone    & $-$1.34 & 13.06 & 12.67 \\
J160659.03-683854.8 & 241.74597 & $-$68.64855 & 0.092009 & 0.06 & Overtone    & $-$0.72 & 12.49 & 12.16 \\
J073817.13-672423.9 & 114.57136 & $-$67.40663 & 0.055886 & 0.08 & Overtone    & $-$0.96 & 13.18 & 12.82 \\
J170344.20-615941.2 & 255.93415 & $-$61.99479 & 0.076917 & 0.07 & Overtone    & $-$0.95 & 13.33 & 12.90  \\
J112420.16-442718.4 & 171.08401 & $-$44.45510  & 0.046792 & 0.07 & Overtone    & $-$1.17 & 12.46 & 12.23 \\
J071855.62-434247.3 & 109.73177 & $-$43.71313 & 0.052369 & 0.09 & Overtone    & $-$1.39  & 11.71 & 11.29 \\
J101138.01-432812.6 & 152.90838 & $-$43.47016 & 0.067405 & 0.07 & Overtone    & $-$1.12 & 11.58 & 11.19 \\
J133601.96-373916.3 & 204.00815 & $-$37.65452 & 0.097528 & 0.06 & Overtone    & $-$0.80  & 12.49 & 12.11 \\
J190405.97-341336.7 & 286.02488 & $-$34.22687 & 0.108644 & 0.16 & Overtone    & $-$0.73 & 13.34 & 12.97 \\
J095042.76-830700.7 & 147.67815 & $-$83.11686 & 0.123165 & 0.08 & Fundamental & 0.16  & 12.17 & 11.69 \\
J082236.01-811518.3 & 125.65004 & $-$81.25508 & 0.077246 & 0.40  & Fundamental & $-$0.24 & 15.19 & 14.77 \\
J173609.95-753309.6 & 264.04146 & $-$75.55266 & 0.048927 & 0.23 & Fundamental & $-$0.04 & 12.59 & 12.32 \\
J174754.43-590251.4 & 266.97681 & $-$59.04760  & 0.082942 & 0.34 & Fundamental & $-$0.39  & 14.39 & 13.95 \\
J115432.36-520615.0 & 178.63482 & $-$52.10417 & 0.174102 & 0.23 & Fundamental & 0.20  & 13.61 & 13.19 \\
J161700.24-451103.3 & 244.25101 & $-$45.18424 & 0.054598 & 0.16 & Fundamental & $-$0.07 & 13.15 & 12.22 \\
J053345.50-414801.1 & 83.43960   & $-$41.80030  & 0.090756 & 0.15 & Fundamental & $-$0.04 & 13.90 & 13.56 \\
J093415.00-323455.0 & 143.56249 & $-$32.58194 & 0.052528 & 0.36 & Fundamental & 0.07  & 12.37 & 12.14 \\
J061658.98-213318.9 & 94.24576  & $-$21.55526 & 0.101825 & 0.40  & Fundamental & $-$0.04 & 13.55 & 13.28 \\
J082639.28-102402.9 & 126.66367 & $-$10.40081 & 0.061929 & 0.20  & Fundamental & $-$0.24 & 12.72 & 12.42 \\
\hline
\end{tabular}
\end{table*}

We used the \verb"Period04" software package \citep{2005CoAst.146...53L} to calculate the Discrete Fourier Transform (DFT) of the \textit{TESS} light curves. Each light curve was iteratively whitened until all the frequencies with signal-to-noise ratios ($\rm SNRs$) $>5$ were retrieved. We have summarized the results of the frequency analysis in Table \ref{tab:finfof} (fundamental mode pulsators) and Table \ref{tab:finfoot} (overtone pulsators). As was noted earlier, the pulsational amplitudes of the fundamental mode pulsators are larger than those of the overtone pulsators. In all but two cases, the fundamental mode is the dominant pulsation mode of the fundamental mode sources, and vice versa. However, the vast majority of these sources really are multi-mode pulsators.

The remarkable light curve of the source ASASSN-V J061658.98$-$213318.9 is composed of two dominant pulsation modes $f_0=9.820608 \rm d^{-1}$ and $f_1=12.698695 \rm d^{-1}$ with a period ratio $P_1/P_0=0.773$ that suggests that $f_1$ is the first overtone. There are also a remarkable array of high amplitude beat frequencies detected at $f_1-f_0$, $f_1+f_0$, $2f_0$, $2f_0-f_1$, $2f_1-f_0$, and $3f_1-2f_0$. The last beat frequency was nominally below our SNR threshold of SNR=5.

\begin{figure*}
	\includegraphics[width=\textwidth]{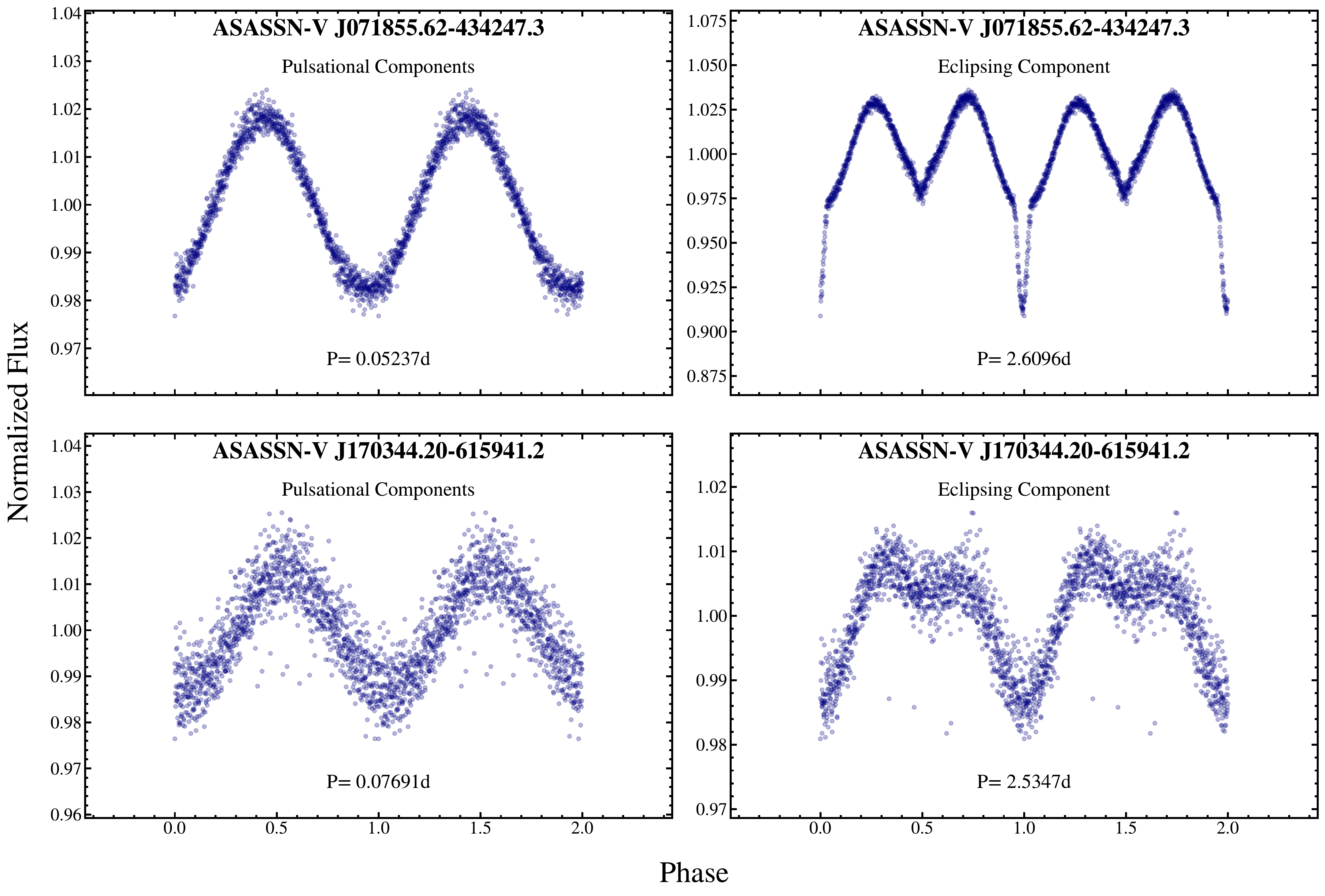}
    \caption{The pulsational (left) and eclipsing (right) components for the two eclipsing binaries ASASSN-V J071855.62$-$434247.3 and ASASSN-V J170344.20$-$615941.2 with a $\delta$ Scuti component. The format is the same as Figure \ref{fig:fig23}.}
    \label{fig:fig25}
\end{figure*}

We also identify two new $\delta$ Scuti eclipsing binaries, ASASSN-V J071855.62$-$434247.3 and ASASSN-V J170344.20$-$615941.2 were both in the sample of overtone pulsators and have short orbital periods of $\rm P_{orb}=2.6096$ d and $\rm P_{orb}=2.5347$ d respectively. Figure \ref{fig:fig25} separately shows the pulsational and eclipsing components of these two systems. We retrieved numerous orbital harmonics with SNR below our threshold (up to $N=14$) from the light curve of ASASSN-V J071855.62$-$434247.3, which were used to construct the eclipsing component shown in Figure \ref{fig:fig25}. The eclipsing component of ASASSN-V J071855.62$-$434247.3 has a strong ellipsoidal component with sharp eclipses, whereas ASASSN-V J170344.20$-$615941.2 appears to be a semi-detached eclipsing system. Such eclipsing binaries can be powerful tools for probing the internal structure of stars, but only ${\sim}100$ such $\delta$ Scuti eclipsing binaries are currently known \citep{2017MNRAS.470..915K}.

\begin{figure}
	\includegraphics[width=0.5\textwidth]{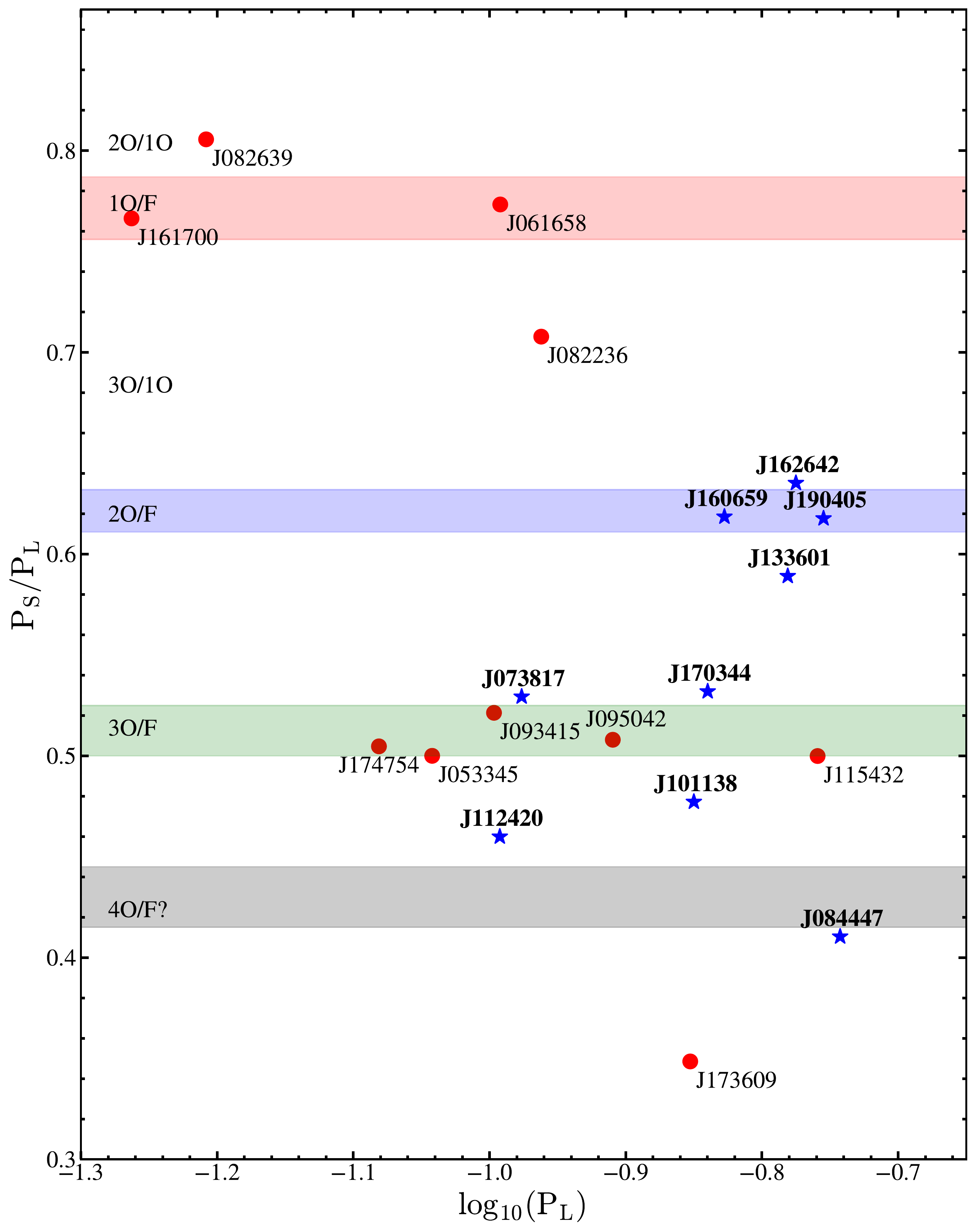}
    \caption{The Petersen diagram for the 10 fundamental mode and 10 overtone $\delta$ Scuti stars where we analyzed the \textit{TESS} light curves. The stars are identified through an abbreviated form of the ASAS-SN IDs listed in Table \ref{tab:srcinfo}. Expected period ratios for combinations of the fundamental mode and the first three overtone modes are annotated \citep{1979ApJ...227..935S}. The range of expected period ratios for $\delta$ Scuti stars pulsating in the fundamental mode and the first overtone (red), the second overtone (blue) or the third overtone (green) are shaded. The likely range of period ratios for the fourth overtone from Figure \ref{fig:fig7} is shaded in black. }
    \label{fig:fig26}
\end{figure}

Figure \ref{fig:fig26} shows a Petersen diagram \citep{1973A&A....27...89P} comparing the longer, fundamental period $P_L$ to the period ratio $P_S/P_L$ with a secondary, shorter period $P_S$. For the fundamental mode stars, we define $P_S$ to be the highest amplitude higher frequency period. For the overtone stars, we were only able to clearly identify the fundamental mode from the light curves for two sources,  ASASSN-V J160659.03$-$683854.8 and ASASSN-V J101138.01$-$432812.6. For the other overtone stars, we estimated the fundamental period $P_L$ using the fundamental mode PLR and the $W_{JK}$ magnitudes of these sources. The source ASASSN-V J071855.62$-$434247.3 is not included here, as the frequency analysis revealed only a single pulsation frequency after the removal of the binary orbital harmonics.

Expected period ratios for combinations of the fundamental mode and the first three overtone modes are highlighted \citep{1979ApJ...227..935S}. Most of the overtone sources are clustered towards period ratios indicative of the second and third overtones. The overtone source ASASSN-V J084447.39$-$724126.6 appears to have $P_S/P_L{\sim}0.42$, which is close to the possible period ratio for the fourth overtone identified in Figure \ref{fig:fig7}.  Half of the fundamental mode sources appear to be multi-mode pulsators pulsating in the third overtone. The fundamental mode source ASASSN-V J173609.95$-$753309.6 has a very unusual period ratio of $P_S/P_L{\sim}0.349$, which suggests that the frequency identified with $f_1=7.122998 \rm d^{-1}$ as $P_S$ is a non-radial pulsation mode.

\begin{table*}
	\centering
	\caption{Pulsation frequencies, amplitudes and SNRs of the 10 fundamental mode $\delta$ Scuti stars in the Southern Hemisphere with \textit{TESS} data}
	\label{tab:finfof}
    \begin{threeparttable}	
\begin{tabular}{rrrrr}
		\hline
		 ASAS-SN ID & Frequency ($\rm d^{-1}$) & Amplitude (ppt) & SNR & Comment \\
		\hline		 
J093415.00-323455.0 & &  & &\\
& 19.037813$\pm0.000011$ & 95.07 $\pm0.23$& 19.4 & \\
& 9.925218$\pm0.000158$ & 6.33$\pm0.09$ & 5.7 & \\

J082639.28-102402.9 & &  & &\\
& 16.147136$\pm0.000047$ & 54.76$\pm0.48$ & 15.6 & \\
& 20.042744$\pm0.001156$ & 2.25$\pm0.16$ & 5.6 & \\

J174754.43-590251.4 & &  & &\\
& 12.056074$\pm0.000018$ & 86.75$\pm0.67$ & 26.1 & \\
& 23.887109$\pm0.000066$ & 24.21$\pm0.17$ & 16.3 & \\

J061658.98-213318.9 & &  & &\\
& 9.820608$\pm0.000042$ & 68.54$\pm0.15$ & 14.5 & $f_0$\\
& 12.698695$\pm0.000044$ & 63.64$\pm1.25$ & 14.3 & $f_1$\\
& 2.878137$\pm0.000108$ & 26.27$\pm0.13$ & 13.7 & $f_1-f_0$\\
& 19.641166$\pm0.000253$ & 11.26$\pm0.13$ & 10.7 &  $2f_0$\\
& 22.517004$\pm0.000114$ & 19.14$\pm1.10$ & 10.6 & $f_1+f_0$\\
& 6.942471$\pm0.000418$ & 6.81$\pm0.13$ & 7.7 & $2f_0-f_1$\\
& 15.579131$\pm0.000504$ & 5.65$\pm0.13$ & 6.6 & $2f_1-f_0$ \\
& 22.602061$\pm0.000289$ & 10.84$\pm0.15$ & 6.0 & \\
& 15.659590$\pm0.000829$ & 3.42$\pm0.13$ & 4.0\tnote{a} & \\
& 18.533129$\pm0.001851$ & 1.53$\pm0.14$ & 2.5\tnote{a} & $3f_1-2f_0$\\

J082236.01-811518.3 & &  & &\\
& 12.945750$\pm0.000015$ & 96.97$\pm1.12$ & 25.5 & \\
& 22.108528$\pm0.000063$ & 22.94$\pm0.23$ & 22.4 & \\
& 9.163372$\pm0.000532$ & 2.73$\pm0.26$ & 6.1 & \\

J095042.76-830700.7  & &  & &\\
& 8.119214$\pm0.000038$ & 17.35$\pm0.12$ & 18.5 & \\
& 15.983336$\pm0.000203$ & 3.24$\pm0.11$ & 14.2 & \\
& 13.386403$\pm0.000323$ & 2.04$\pm0.10$ & 13.8 & \\
& 8.430682$\pm0.000079$ & 8.37$\pm0.13$ & 9.0 & \\
& 15.608756$\pm0.000323$ & 2.03$\pm0.11$ & 8.4 & \\
& 8.056102$\pm0.000130$ & 5.07$\pm0.12$ & 5.4 & \\

J115432.36-520615.0  & &  & &\\
& 5.743287$\pm0.000040$ & 75.77$\pm0.81$ & 16.1 & \\
& 11.488557$\pm0.000224$ & 13.35$\pm0.15$ & 11.0 & \\
& 7.622044$\pm0.000428$ & 7.00$\pm0.16$ & 6.6 & \\

J053345.50-414801.1 & &  & &\\
& 11.018842$\pm0.000036$ & 36.82$\pm0.14$ & 21.5 & $f_0$\\
& 22.037130$\pm0.000456$ & 2.91$\pm0.12$ & 12.0 & $2f_0$\\
& 18.254843$\pm0.000553$ & 2.40$\pm0.14$ & 11.0 & \\

J173609.95-753309.6  & &  & &\\
& 20.438721$\pm0.000011$ & 51.32$\pm0.07$ & 20.3 & \\
& 7.122998$\pm0.000480$ & 1.21$\pm0.05$ & 7.9 & NR?\tnote{b}\\

J161700.24-451103.3  & &  & &\\
& 18.316320$\pm0.000076$ & 28.59$\pm0.22$ & 14.9 & \\
& 23.897836$\pm0.000806$ & 2.68$\pm0.10$ & 8.3 & \\

		\hline
\hline
\end{tabular}
\begin{tablenotes}
\item[a] {Detections with SNR<5}
\item[b] {Possible non-radial mode}
\end{tablenotes}
\end{threeparttable}   
\end{table*}

\begin{table*}
	\centering
	\caption{Pulsation frequencies, amplitudes and SNRs of the 10 overtone $\delta$ Scuti stars in the Southern Hemisphere with \textit{TESS} data}
	\label{tab:finfoot}
\begin{tabular}{rrrrr}
		\hline
		 ASAS-SN ID & Frequency ($\rm d^{-1}$) & Amplitude (ppt) & SNR & Comment \\
		\hline		 
J073817.13-672423.9 & &  & &\\
& 17.893421$\pm0.000006$ & 9.75$\pm0.05$ & 55.0 & \\
& 18.501227$\pm0.000011$ & 5.49$\pm0.04$ & 31.7 & \\

J160659.03-683854.8 & &  & &\\
& 10.868886$\pm0.000113$ & 11.61$\pm0.08$ & 13.0 & \\
& 9.508483$\pm0.000257$ & 5.09$\pm0.09$ & 12.4 & \\
& 11.630640$\pm0.000195$ & 6.71$\pm0.08$ & 7.5 & \\
& 6.744660$\pm0.001870$ & 0.70$\pm0.07$ & 6.9 & \\

J133601.96-373916.3 & &  & &\\
& 10.253547$\pm0.000151$ & 11.12$\pm0.09$ & 12.8 & \\
& 7.586851$\pm0.000430$ & 3.91$\pm0.16$ & 7.9 & \\

J162642.94-724540.2 & &  & &\\
& 9.378243$\pm0.000124$ & 21.69$\pm0.11$ & 9.3 & \\
& 12.036304$\pm0.000165$ & 16.24$\pm0.10$ & 9.2 & \\
& 17.812164$\pm0.001497$ & 1.79$\pm0.10$ & 5.7 & \\
& 21.122983$\pm0.001309$ & 2.05$\pm0.11$ & 5.5 & \\

J112420.16-442718.4 & &  & &\\
& 21.369170$\pm0.000167$ & 10.23$\pm0.23$ & 13.7 & \\
& 18.381295$\pm0.000518$ & 3.29$\pm0.08$ & 8.6 & \\
& 16.744625$\pm0.000914$ & 1.87$\pm0.08$ & 8.4 & \\

J101138.01-432812.6 & &  & &\\
& 14.835528$\pm0.000235$ & 11.03$\pm0.14$ & 14.5 & \\
& 7.242484$\pm0.000564$ & 4.61$\pm0.14$ & 9.4 & \\
& 12.390962$\pm0.000346$ & 7.50$\pm0.13$ & 8.1 & \\
& 11.837701$\pm0.000386$ & 6.72$\pm0.14$ & 7.5 & \\
& 7.088432$\pm0.000991$ & 2.62$\pm0.13$ & 5.4 & \\

J190405.97-341336.7 & &  & &\\
& 9.204886$\pm0.000110$ & 45.52$\pm0.29$ & 15.4 & \\
& 14.201801$\pm0.000759$ & 6.58$\pm0.25$ & 10.3 & \\
& 11.958898$\pm0.000873$ & 5.72$\pm0.28$ & 9.1  & \\
& 18.407353$\pm0.001206$ & 4.14$\pm0.29$ & 7.6 & \\
& 6.190006$\pm0.001103$ & 4.53$\pm0.26$ & 5.5 & \\

J084447.39-724126.6 & &  & &\\
& 13.471703$\pm0.000030$ & 7.95$\pm0.05$ & 23.6 & \\
& 12.146557$\pm0.000070$ & 3.35$\pm0.05$ & 16.8 & \\
& 12.964131$\pm0.000054$ & 4.35$\pm0.06$ & 12.4 & \\
& 15.226689$\pm0.000209$ & 1.13$\pm0.05$ & 10.5 & \\

J071855.62-434247.3 & &  & & \\
& 19.095246$\pm0.000102$ & 18.16$\pm0.10$ & 15.3 & \\
& 0.765662$\pm0.000059$ & 31.55$\pm0.10$ & 11.7 & $\rm P_{orb}/2$, $\rm P_{orb}=2.6096$ d \\

J170344.20-615941.2 & &  & & \\
& 13.001638$\pm0.000199$ & 12.82$\pm0.21$ & 15.4 & \\
& 23.413779$\pm0.001242$ & 2.06$\pm0.14$ & 7.9 & \\	
& 0.394516$\pm0.000274$ & 9.32$\pm0.14$ & 6.5 & $\rm P_{orb}$, $\rm P_{orb}=2.5347$ d \\
		\hline
\hline
\end{tabular}
\end{table*}

\section{Conclusions}

We analyzed the all-sky catalogue of 8418 $\delta$ Scuti stars in the ASAS-SN V-band catalog of variables which includes 3322 new discoveries. We derive period-luminosity relationships (PLRs) for fundamental mode and overtone $\delta$ Scuti stars in the $W_{JK}$, $V$, Gaia DR2 $G$, $J$, $H$, $K_s$ and $W_1$ bands.
We identified a peak in the distribution of $\Delta \log \rm P$ from the fundamental PLR fit corresponding to a period ratio of $P_{4O}/P_F{\sim}0.43$, which is likely to be the fourth overtone. We encourage a theoretical study of the fourth overtone mode in $\delta$ Scuti stars, as the empirical evidence points towards a population of $\delta$ Scuti stars pulsating with a dominant fourth overtone. We also investigated how the PLRs vary with the pulsational amplitude and Gaia DR2 distance, but did not find a significant variation in the PLRs with either parameter. The PLRs have shortcomings because the ASAS-SN photometric data and the period are not adequate to clearly separate all the mode sequences. This would be more feasible given a complete analysis of the \textit{TESS} light curves.

We also cross-matched the catalogue to wide-field spectroscopic surveys, which resulted in 972 cross-matches. For the stars with spectroscopic parameters, we find that $\rm T_{eff}$, and $\rm \log (g)$ are as expected from theoretical models. We used the calibrations in \citet{2010A&ARv..18...67T} to derive masses and radii for these sources to find that the distributions of the radii of the fundamental mode and overtone pulsators were largely the same, but the distributions of masses were somewhat different, with an excess of overtone pulsators with masses $M>1.8M_\odot$ compared to the fundamental mode pulsators. An analysis of the pulsational constants ($Q$) for overtone pulsators in the spectroscopic sub-sample shows multiple peaks corresponding to the first, second and third overtones visible in the data, and an absence of sources with $Q$ close to the fundamental mode. This is in contrast to the fundamental mode pulsators that have $Q$ values close to the expected value for the radial fundamental mode, and the first overtone. We again noted evidence for sources pulsating in the fourth overtone.

The periods of the fundamental mode $\delta$ Scuti stars vary with the metallicity, from $\log _{10} (\rm P/days){\sim}-1.1$ for sources with $\rm [Fe/H]<-0.3$ to $\log _{10} (\rm P/days){\sim}-0.9$ for sources with super-solar metallicities $\rm [Fe/H]>0$. This leads to a period gradient with distance from the Galactic mid-plane. We find that $\delta$ Scuti stars with periods $\rm P>0.100\,d$ are predominantly located towards the Galactic disk ($\rm |Z|<0.5\,kpc$).  The median period at a scale height of $\rm Z{\sim}0\, kpc$ increases with the Galactocentric radius $\rm R$, from $\log _{10} (\rm P){\sim}-0.94$ for sources with $\rm R>9\, kpc$ to $\log _{10} (\rm P){\sim}-0.85$ for sources with $\rm R<7\, kpc$, which is indicative of a radial metallicity gradient.

As an experiment, we examined \textit{TESS} light curves for a sample of 10 fundamental mode and 10 overtone $\delta$ Scuti stars discovered by ASAS-SN. A frequency analysis of these light curves showed that 19 of the 20 sources were multi-mode pulsators. Overtone sources had a dominant overtone mode and the expected fundamental mode was recovered in only two light curves. The analysis of the Petersen diagram showed that most of the overtone sources were clustered towards period ratios indicative of the second and third overtones, which is consistent with the analysis of the pulsational constants. We find that the unusual light curve of the source ASASSN-V J061658.98$-$213318.9 is composed of two dominant pulsation modes $f_0=9.820608 \rm d^{-1}$ and $f_1=12.698695 \rm d^{-1}$ with a period ratio $P_1/P_0=0.773$ that suggests that $f_1$ is the first overtone, and a remarkable array of high amplitude beat frequencies. We also identified two new $\delta$ Scuti eclipsing binaries, ASASSN-V J071855.62$-$434247.3 and ASASSN-V J170344.20$-$615941.2, which were both in the sample of overtone pulsators, and have short orbital periods of $\rm P_{orb}=2.6096$ d and $\rm P_{orb}=2.5347$ d respectively.

\section*{Acknowledgements}

We thank the anonymous referee for the very useful comments that improved our presentation of this work. We thank Dr. Jennifer van Saders, Dr. Radoslaw Poleski and Dr. Marc Pinsonneault for useful discussions on this manuscript. We thank the Las Cumbres Observatory and its staff for its continuing support of the ASAS-SN project. We also thank the Ohio State University College of Arts and Sciences Technology Services for helping us set up and maintain the ASAS-SN variable stars and photometry databases.

ASAS-SN is supported by the Gordon and Betty Moore
Foundation through grant GBMF5490 to the Ohio State
University, and NSF grants AST-1515927 and AST-1908570. Development of
ASAS-SN has been supported by NSF grant AST-0908816,
the Mt. Cuba Astronomical Foundation, the Center for Cosmology 
and AstroParticle Physics at the Ohio State University, 
the Chinese Academy of Sciences South America Center
for Astronomy (CAS- SACA), the Villum Foundation, and
George Skestos. 

KZS and CSK are supported by NSF grants AST-1515927, AST-1814440, and AST-1908570. This work is supported in part by Scialog Scholar grant 24216 from the Research Corporation. PJV is supported by the National Science Foundation Graduate Research Fellowship Program Under Grant No. DGE-1343012. B.J.S is supported by NSF grants AST-1908952, AST-1920392, and AST-1911074. TAT acknowledges support from a Simons Foundation Fellowship and from an IBM Einstein Fellowship from the Institute for Advanced Study, Princeton. Support for JLP is provided in part by the Ministry of Economy, Development, and Tourism's Millennium Science Initiative through grant IC120009, awarded to The Millennium Institute of Astrophysics, MAS. Support for OP has been provided by INTER-EXCELLENCE grant LTAUSA18093 from the Czech Ministry of Education, Youth, and Sports. The research of OP has also been supported by Horizon 2020 ERC Starting Grant ``Cat-In-hAT'' (grant agreement \#803158) and PRIMUS/SCI/17 award from Charles University. This work was partly supported by NSFC 11721303.

This paper includes data collected by the TESS mission, which are publicly available from the Mikulski Archive for Space Telescopes (MAST). Funding for the TESS mission is provided by NASA's Science Mission directorate.

 We thank Ethan Kruse for uploading the TESS FFIs to YouTube, as these videos were invaluable when investigating the prevalence of scattered light artifacts for our targets.

This work has made use of data from the European Space Agency (ESA)
mission {\it Gaia} (\url{https://www.cosmos.esa.int/gaia}), processed by
the {\it Gaia} Data Processing and Analysis Consortium. This publication makes 
use of data products from the Two Micron All Sky Survey, as well as
data products from the Wide-field Infrared Survey Explorer.
This research was also made possible through the use of the AAVSO Photometric 
All-Sky Survey (APASS), funded by the Robert Martin Ayers Sciences Fund. 

This research has made use of the VizieR catalogue access tool, CDS, Strasbourg, France. 
This research also made use of Astropy, a community-developed core Python package for 
Astronomy (Astropy Collaboration, 2013).







\bsp	
\label{lastpage}
\end{document}